\documentclass{aa}  

\usepackage{graphicx}
\usepackage{xcolor}
\usepackage{subfig}
\usepackage{txfonts}
\newcommand{\kepler}{\emph{Kepler}\xspace}
\newcommand{\gaia}{\emph{Gaia}\xspace}

\newcommand{\figref}[1]{Fig.~\ref{#1}}

\newcommand{\teff}{\ensuremath{T_{\textup{eff}}}\xspace}
\newcommand{\logg}{\ensuremath{\log g}\xspace}

\newcommand{\flag}[1]{\texttt{#1}}
\begin{document}

   \title{Time evolution of Ce as traced by APOGEE using giant stars observed with the {\it Kepler}, TESS and K2 missions}


   \author{G. Casali\inst{\ref{difa},\ref{oas}} \and
          V. Grisoni\inst{\ref{difa},\ref{oas}} \and
          A. Miglio\inst{\ref{difa},\ref{oas}} \and
          C. Chiappini\inst{\ref{aip}} \and
          M. Matteuzzi\inst{\ref{difa},\ref{oas}} \and
          L. Magrini\inst{\ref{oaa}} \and
          E. Willett\inst{\ref{birm}} \and
          G. Cescutti\inst{\ref{units},\ref{oat},\ref{infnts}} \and
          F. Matteucci\inst{\ref{units},\ref{oat},\ref{infnts},\ref{ifpu}} \and
          A. Stokholm\inst{\ref{difa},\ref{oas}}
          \and
          M. Tailo\inst{\ref{difa}}
          \and
          J. Montalb\'an\inst{\ref{difa}} \and
          Y. Elsworth\inst{\ref{birm}} \and
          B. Mosser\inst{\ref{lesia}}
          }

\institute{
{Dipartimento di Fisica e Astronomia, Universit\`a di Bologna, Via Gobetti 93/2, I-40129 Bologna, Italy}\label{difa}\\
\email{giada.casali@inaf.it}
\and
{INAF – Osservatorio di Astrofisica e Scienza dello Spazio, Via P. Gobetti 93/3, 40129 Bologna, Italy} \label{oas} 
\and
{Leibniz-Institut fur Astrophysik Potsdam (AIP), An der Sternwarte 16, D-14482 Potsdam, Germany} \label{aip}
\and
{INAF – Osservatorio Astrofisico di Arcetri, Largo E. Fermi 5, 50125 Firenze, Italy} \label{oaa}
\and
{School of Physics \& Astronomy, University of Birmingham, Edgbaston, Birmingham, B15 2TT, UK} \label{birm}
\and
{Dipartimento di Fisica, Sezione di Astronomia, Universit\`a di Trieste, Via G. B. Tiepolo 11, 34143 Trieste, Italy} \label{units}
\and
{INAF – Osservatorio Astronomico di Trieste, Via Tiepolo 11, 34143, Trieste, Italy} \label{oat}
\and
{INFN – Sezione di Trieste, Via A. Valerio 2, 34127 Trieste, Italy} \label{infnts}
\and
{IFPU, Institute for the Fundamental Physics of the Universe, Via Beirut, 2, I-34151 Trieste, Italy} \label{ifpu}
\and
{LESIA, Observatoire de Paris, Universit\'e PSL, CNRS, Sorbonne Universit\'e, Universit\'e de Paris, 92195 Meudon, France} \label{lesia}
}

 
  \abstract
   {Abundances of slow neutron-capture process elements in stars with exquisite asteroseismic, spectroscopic, and astrometric constraints offer a novel opportunity to study stellar evolution, nucleosynthesis, and Galactic chemical evolution. }
   {We aim to investigate one of the least studied s-process elements in the literature, cerium (Ce), using stars with asteroseismic constraints from the {\it Kepler}, K2 and TESS missions. }
   {We combine the global asteroseismic parameters derived from precise light curves obtained by the {\it Kepler}, K2 and TESS missions with stellar parameters and chemical abundances from the latest data release of the large spectroscopic survey APOGEE and astrometric data from the Gaia mission. Finally, we compute stellar ages using the code PARAM with a Bayesian estimation method.}
   {We investigate the different trends of [Ce/Fe] as a function of metallicity, [$\alpha$/Fe] and age taking into account the dependence on the radial position, specially in the case of K2 targets which cover a large Galactocentric range. We, finally, explore the [Ce/$\alpha$] ratios as a function of age in different Galactocentric intervals.}
   {The studied trends display a strong dependence of the Ce abundances on the metallicity and star formation history. Indeed, the [Ce/Fe] ratio shows a non-monotonic dependence on [Fe/H] with a peak around $-0.2$ dex. Moreover, younger stars have higher [Ce/Fe] and [Ce/$\alpha$] ratios than older stars, confirming the latest contribution of low- and intermediate-mass asymptotic giant branch stars to the Galactic chemical enrichment. In addition, the trends of [Ce/Fe] and [Ce/$\alpha$] with age become steeper moving towards the outer regions of the Galactic disc, demonstrating a more intense star formation in the inner regions than in the outer regions. Ce is thus a potentially interesting element to help constraining stellar yields and the inside-out formation of the Milky Way disc. However, the large scatter in all the relations studied here, suggests that spectroscopic uncertainties for this element are still too large.}

   \keywords{Galaxy: evolution -- Galaxy: abundances -- Galaxy: disk -- stars: abundances -- stars: late-type –asteroseismology}

   \maketitle
%

\section{Introduction}

Our Universe is enriched in different chemical elements on different timescales (due to their different nucleosynthetic origin). This suggests that some chemical abundance ratios can be used to trace different star formation histories. One classical abundance ratio often used in Galactic Archaeology is the [$\alpha$/Fe] ratio \citep[e.g.,][]{Matteucci2021}. Other abundance ratios are also expected to vary with age, as for instance the ratio between slow- (s-) process and $\alpha$ elements -- the so called chemical clocks. Recent data on open clusters \citep[e.g.,][]{casamiquela21,viscasillas22}, and on field stars \citep[e.g.,][]{spina18,nissen20,Morel21} for which it was possible to measure ages, have confirmed the theoretical expectations in general terms.



S-process elements can be produced in massive stars \citep[weak component, 60 $<$ A $<$ 90, with A being the atomic mass number,][]{pignatari2010} or in low- and intermediate-mass asymptotic giant branch (AGB) stars with a main component \citep[90 $<$ A $<$ 204,][]{lugaro2003} and/or with a strong component \citep[by low-metallicity AGB stars, 204 $<$ A $<$ 209,][]{Gallino1998}. Cerium (Ce) has mostly been produced by the main s-process component \citep[83.5 $\pm$ 5.9\% at solar metallicity,][]{bisterzo2014} in low-mass AGB stars ($1.5-3 M_{\odot}$).

However, the standard view of the s-process in massive stars might be modified in rotating stars because of the rotational mixing operating between the H-shell and He-core during the core helium burning phase. Several observational signatures support an enhancement of s-process elements (up to A $\sim$ 140) in massive rotating stars \citep{pignatari08,chiappini13,cescutti13,cescutti14}. 
Recently, overabundance of Ce was observed in metal-poor bulge stars \citep{razera22}, which was interpreted as a result of an early enrichment by fast rotating massive stars, or spinstars \citep{meynet02,chiappini11,cescutti18}.


In the last decade, [Y/Mg] and [Y/Al] have been among the most studied chemical clocks \citep{dasilva12, nissen15, feltzing17,slumstrup17,spina18,delgado19,anders18,casali20,jofre20,nissen20}. However, recent works demonstrated not only their metallicity dependence, but also the non-universality of the chemical clock-age relations, with variations of the shape of the relations at different Galactocentric distance  \citep{casali20,magrini21,casamiquela21,viscasillas22}.

Owing to the large uncertainties on ages of field stars, these studies have been mostly restricted to clusters. 
In this context, asteroseismology comes to our aid. Through asteroseismology, we are able to detect solar-like pulsations in thousands of G-K giants using data collected by the COnvection ROtation and planetary Transits \citep[CoRoT,][]{baglin06}, {\it Kepler} \citep{gilliland10}, K2 \citep{howell14} and Transiting Exoplanet Survey Satellite \citep[TESS,][]{ricker14} missions. The pulsation frequencies are directly linked to the stellar structure and thus provide tight constraints on stellar properties (radius, mass, age) and evolutionary state \citep{chaplin13}. 

In this paper, we focus on the evolution of one of the least studied s-process elements in literature, Ce, in a sample of field stars for which precise asteroseismic ages are available. 
The [Ce/Fe]-age (and also [Ce/$\alpha$]-age) relation has been studied in open clusters and solar twins by, e.g., \citet{maiorca11}, \citet{spina18}, \citet{magrini18}, \citet{delgado19}, \citet{grazina21}, \citet{casamiquela21}, \citet{viscasillas22}, \citet{salessilva22}. They found an increase of [Ce/Fe] with decreasing stellar age, except for \citet{grazina21} where they found an almost flat trend. 

Here, we complement such studies with field stars sampling a larger age baseline than possible with clusters only, thanks to asteroseismic ages coming from three different missions ({\it Kepler}, TESS and K2).
We will use Ce abundances published in the latest data release (DR17) of the high-resolution spectroscopic survey, Apache Point Observatory Galactic Evolution Experiment, APOGEE \citep{majewski17,apogeedr17}.  
Our aim is to investigate the time evolution of Ce across the Milky Way ({\it Kepler} and TESS are focused on the solar neighbourhood, $7.5 < R_{GC} < 8.5$~kpc, but K2 covers a large range in Galactocentric radii, $4 < R_{GC} < 12$~kpc). With this dataset we investigate the [Ce/Fe] trends as a function of the metallicity [Fe/H] and [$\alpha$/Fe], its time evolution and [Ce/$\alpha$] as chemical clocks.

The paper is structured as follows. In Sect.~\ref{sec:datasample}, we describe the data samples; in Sect. ~\ref{sec:comparison}, we present a comparison between the Ce abundances used in this work and those determined in other surveys;
in Sect.~\ref{sec:cefe} and \ref{sec:tempevol}, we show our results for the [Ce/Fe] trends and the Ce abundance time evolution; in Sect.~\ref{sec:models} we show a quantitative comparison of our results with Galactic chemical evolution models; finally, in Sect.~\ref{sec:conclusions}, we summarise and conclude.

\section{Stellar samples}
\label{sec:datasample}
Our data samples combine the global asteroseismic parameters measured from light curves obtained by the {\it Kepler} \citep{Borucki2010}, K2 \citep{howell14} and TESS \citep{Ricker2015} missions with stellar parameters and chemical abundances inferred from near-infrared high-resolution spectra taken by the APOGEE DR17 survey \citep{apogeedr17}. These targets are then cross-matched with the Early Third Data Release of Gaia \citep[\gaia EDR3,][]{gaiaedr3}.

\subsection{Spectroscopic constraints}
The atmospheric parameters and abundances used in this paper are produced by the standard data analysis pipeline, the APOGEE Stellar Parameters and Chemical Abundances Pipeline \citep[ASPCAP,][]{garcia2016}. A full description of the pipeline is given in Holtzman et al. (in prep) and a description of an earlier implementation of this pipeline can be found in \citet{garcia2016}.

\subsection{Asteroseismic constraints}
We consider three main samples:
\begin{itemize}
    \item Our first data sample is composed by $\sim 5600$ \textit{Kepler} solar-like oscillating giants whose spectroscopic parameters are available from APOGEE DR17 \citep{apogeedr17}. 
    The list of targets and global asteroseismic constraints  corresponds to the reference sample (R1) explored in  \citet{miglio21}, but updated to use photospheric parameters from APOGEE DR17.


\item
The second data sample is composed by data collected by the TESS mission during the first year of observations in its southern continuous viewing zone \citep[SCVZ, see][for more details]{mackereth21}.

The sample, which corresponds to the \emph{gold sample} in \citet{mackereth21} with APOGEE DR17 parameters available, is composed of $\sim 1700$ stars, with $\sim 1$-yr long TESS observations. 

\item
The third dataset is composed by data collected by the K2 mission  in 20 observational campaigns along the ecliptic. Unlike the previous two missions, K2 observed stars in a wide range of Galactocentric radii crucially extending the regions sampled by \textit{Kepler} and TESS. Atmospheric parameters from APOGEE DR17 stars are available for a total of $\sim 11,000$ stars with asteroseismic constraints from K2. 
For this sample we used the global parameter $\nu_{max}$, \citet{Elsworth2020} pipeline,  as the only asteroseismic constraint (see below for more details).

\end{itemize}
\subsection{Inferring stellar masses, ages, and radii}
\label{sec:method}
Masses, radii, and ages are computed using the code PARAM \citep{dasilva2006,rodrigues2017} which makes use of a Bayesian estimation method.
We input a combination of seismic indices and spectroscopic constraints, such as $\nu_{max}$, $\Delta \nu$, [Fe/H], [$\alpha$/Fe], and $T_{\rm eff}$.
A detailed explanation of the method is available in \citet{miglio21}.

For the K2 dataset (80d-long light curves), however, we include as constraint the luminosity, $L$, instead of $\Delta \nu$, because the latter is affected by significant systematic uncertainties in shorter datasets, as described in Willett at al. in prep. (see also \citealt{Tailo2022}). 
The luminosity is computed using the magnitude $K_{s}$ from 2MASS \citep{2mass}, distance deduced from the Gaia parallax, extinction determined using the \texttt{Bayestar2019} map \citep{bayestar19} through the \texttt{dustmaps} Python package \citep{dustmaps}, and bolometric corrections computed  through the code by \citet{casagrande14,casagrande18b,casagrande18a}. In addition, the luminosity takes into account the global median Gaia parallax zero point offset of $-17\mu as$ based on quasars \citep{lindegren21}.

When using PARAM, we consider a lower-limit of 0.05 dex for the uncertainty on [Fe/H] and a limit of 50 K for $T_{\rm eff}$. This choice is due to the very small uncertainties present in APOGEE DR17 (typical uncertainties are $\sigma_{T_{\rm eff}} \sim 10$~K, $\sigma_{\log~g} \sim 0.03$~dex and $\sigma_{{\rm [Fe/H]}} \sim 0.01$~dex) as the quoted uncertainties are internal errors only, and cross-validation against other surveys shows larger systematic differences \citep[e.g., see][]{Rendle2019, Hekker2019, Anguiano2018}. 
Moreover, model-predicted \teff suffer from large uncertainties associated with the modelling of outer boundary conditions and near-surface convection, hence we prefer to downplay the role of $T_{\rm eff}$. The grid of stellar models used in PARAM for this work is the reference grid adopted in \citet{miglio21}, labelled G2 in their work.

\begin{figure}[ht]
\centering
\includegraphics[scale=0.45]{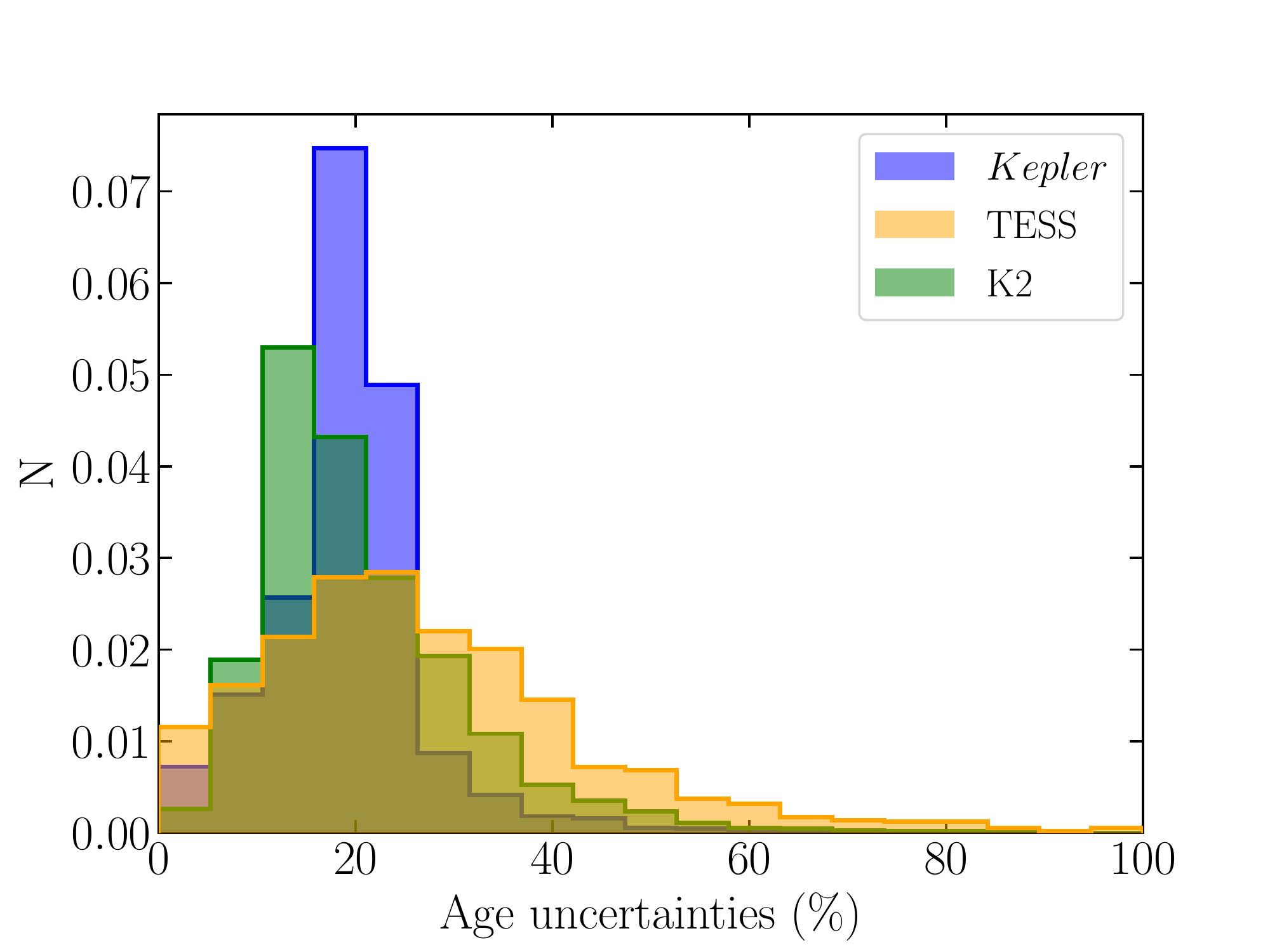}
\caption{Distribution of the age uncertainties for the three data samples. The area under each histogram is integrated to 1. \label{fig:hist_ageunc}}
\end{figure}

The distributions of uncertainties in stellar age for the three samples are shown in Fig.~\ref{fig:hist_ageunc}. While the longer duration of {\it Kepler}'s time series leads to a lower median age uncertainty (20\%) compared to TESS' (28\%), the relatively low median age uncertainty for  K2 (18\%) can be attributed to the use of a very precise (6\%) luminosity instead of  $\Delta \nu$.

For a more detailed description on all datasets, we refer to the work by Willett et al. (in prep).

Finally, to investigate whether the use of $L$ instead $\Delta \nu$ while inferring ages could lead to a significant bias we have compared ages based on $\nu_{max}$ and $L$ with those obtained from $\nu_{max}$ and $\Delta \nu$ for the \kepler and TESS samples (where $\Delta \nu$ is not affected by systematic uncertainties). As a result, we found no significant difference between the two inferred ages for both samples (difference $\lesssim 1 \sigma$).

\subsection{Selection criteria}
\label{quality_check}
In this section we describe how we select stars with robust spectroscopy and age estimates. 
Namely,  we keep stars with $S/N > 100$ and $\sigma(\rm Ce) < 0.2$ dex.
Then, we remove stars with the following flags from the APOGEE survey: \flag{STARFLAG} = \flag{VERY\_BRIGHT\_NEIGHBOR}, \flag{LOW\_SNR}, \flag{PERSIST\_HIGH}, \flag{PERSIST\_JUMP\_POS}, \flag{PERSIST\_JUMP\_NEG}, \flag{SUSPECT\_RV\_COMBINATION} or \flag{ASPCAPFLAG} = \flag{STAR\_BAD}, \flag{STAR\_WARN} and \flag{ELEMFLAG} = \flag{GRIDEDGE\_BAD}, \flag{CALRANGE\_BAD}, \flag{OTHER\_BAD}, \flag{FERRE\_FAIL} for Ce.

We also restrict the sample to stars with estimated radii smaller than 11 $R_{\odot}$ \cite[following the same cut present in][]{miglio21}. This avoids the contamination by early-AGB stars and removes stars with relatively low $\nu_{max}$, a domain where seismic inferences have not been extensively tested so far. 
Moreover, we compute the distribution of age uncertainties over the whole population, and remove stars whose uncertainties lie in the top 5\%. 

Finally, the K2 sample contains very metal poor stars with respect to those of {\it Kepler} and TESS. In order to have samples in the same range of metallicity, we select stars in $\rm -1 \leq [Fe/H] < 0.4$~dex. 


This reduces the initial {\it Kepler} sample of $\sim 5600$ stars to $\sim 2700$, the TESS sample of $\sim 1700$ stars to $\sim 900$ and the K2 sample of $\sim 11,000$ stars to $\sim 4000$. 

\section{Comparison with the literature}
\label{sec:comparison}
In order to validate the Ce abundances we will use in this study, we compare the stars in common between APOGEE DR17 (used in this work) and the optical counterpart from the Gaia-ESO DR5 survey \citep{randich22}. Moreover, we also compare the Ce abundances measured from APOGEE spectra using three different pipelines: ASPCAP, the standard pipeline used to get the abundances in the APOGEE DR17 survey, BACCHUS \citep[Brussels Automatic Code for Characterizing High Accuracy Spectra,][]{masseron16} used in \citet{salessilva22}, and BAWLAS (BACCHUS Analysis of Weak Lines in APOGEE Spectra) used in \cite{hayes2022}.
\citet{salessilva22} redetermined the Ce abundances from the measurements of Ce II lines in the APOGEE DR16 spectra of the member stars of the open clusters present in the Open Cluster Chemical Abundances and Mapping (OCCAM) sample \citep{donor2020}, using the BACCHUS pipeline. 
\citet{hayes2022}, instead, reanalysed $\sim 120,000$ APOGEE DR17 spectra with the updated version of the BACCHUS for weak and blended lines, BAWLAS.

\begin{figure}[ht]
\centering
\includegraphics[scale=0.3]{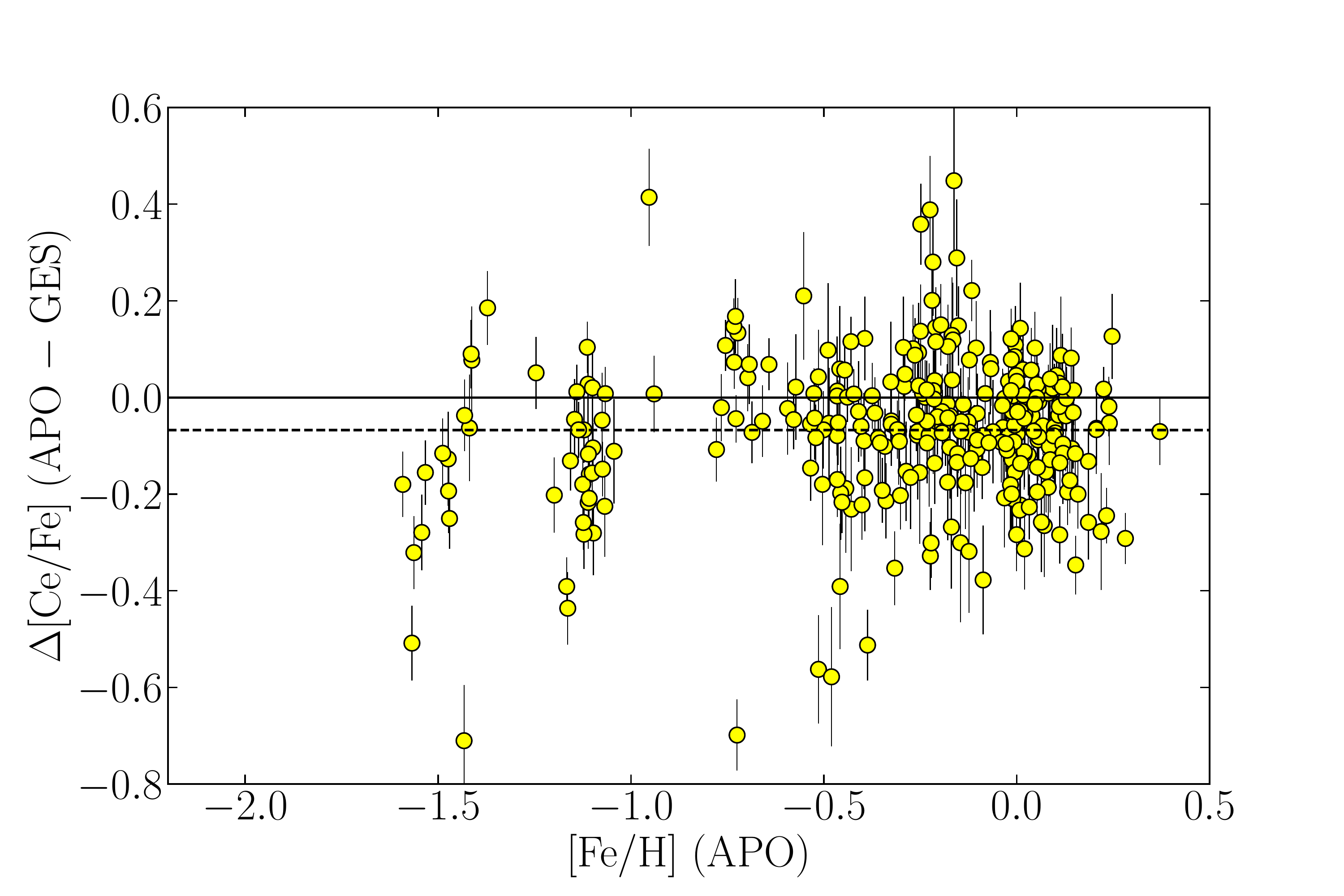}
\caption{Comparison between the [Ce/Fe] ratio using the abundances in the APOGEE DR17 (this work) and the Gaia-ESO abundances \citep{randich22}. The dashed line indicates the median, whereas the solid line indicates the zero difference.  \label{fig:comp1}}
\end{figure}

For the APOGEE DR17 sample, we select stars with high quality Ce abundances (same selection in $\rm \sigma(Ce)$, $S/N$, and APOGEE flags shown in Sect.~\ref{quality_check}).
Instead, for Gaia-ESO, we keep stars with Ce measurements from at least two lines in their spectra.
In \figref{fig:comp1}, we compare the 298 stars in common between APOGEE DR17 with Gaia-ESO DR5. The mean difference in [Ce/Fe] is $\rm \Delta [Ce/Fe] (APOGEE - GES) = -0.07 \pm 0.15$~dex. The scatter is larger at lower metallicity (see Fig.~\ref{fig:comp1}).

\begin{figure}[ht]
\centering
\includegraphics[scale=0.3]{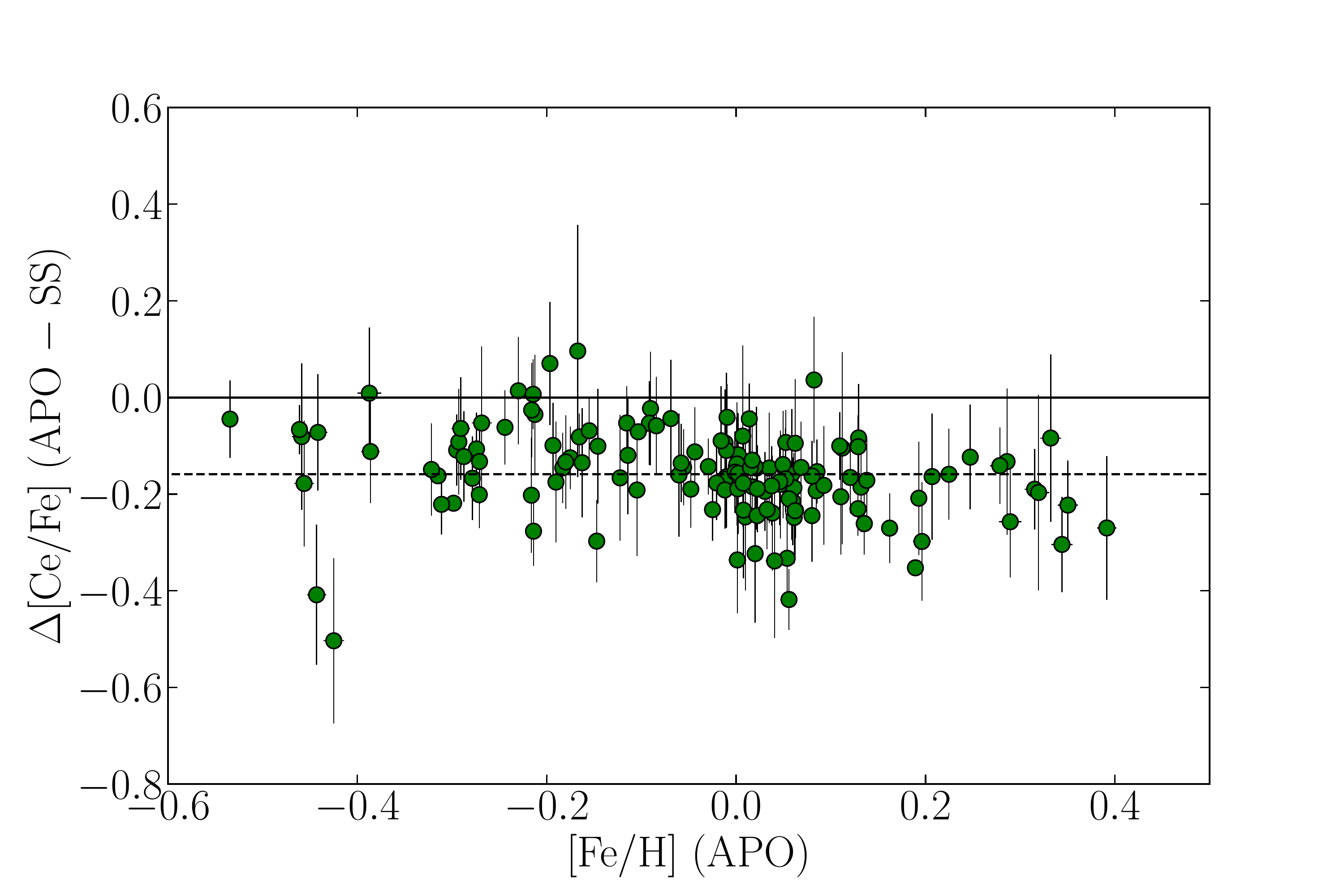}
\caption{Comparison between the [Ce/Fe] ratio using the abundances in the APOGEE DR17 (this work) and the abundances from \citet{salessilva22}.  The dashed line is the median, whereas the solid line indicates the zero difference. \label{fig:comp2}}
\end{figure}

Comparing, instead, the APOGEE DR17 with \citet{salessilva22}, we find 162 member stars in common. They have a mean difference 
$\rm \Delta [Ce/Fe] (APOGEE - SS)$ of $-0.13 \pm 0.09$ dex (see Fig.~\ref{fig:comp2}). 

\begin{figure}[ht]
\centering
\includegraphics[scale=0.3]{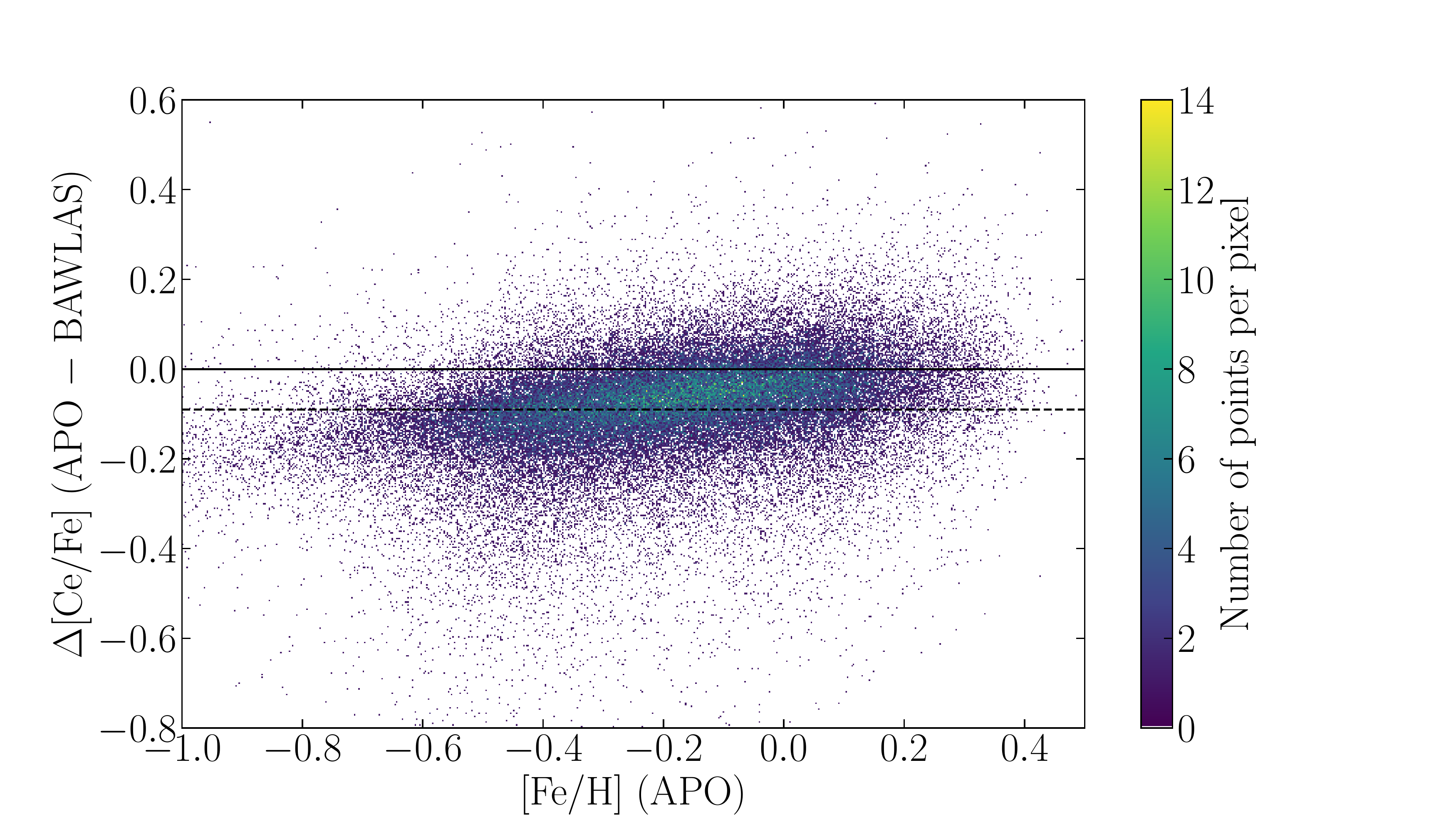}
\caption{Comparison between the [Ce/Fe] ratio using the abundances in the APOGEE DR17 (this work) and the BAWLAS abundances \citep{hayes2022}.  The dashed line is the median, whereas the solid line indicates the zero difference.  \label{fig:comp3}}
\end{figure}

Finally, the comparison between APOGEE DR17 and \citet{hayes2022} shows a difference of $\rm \Delta [Ce/Fe] (APOGEE - BAWLAS) = -0.09 \pm 0.13$ dex. 
The chemical abundance patterns of Ce from ASPCAP and BAWLAS are relatively similar. However, \citet{hayes2022} find a trend in the differences between the two Ce measurements as a function of [Fe/H] at lower metallicities. Figure~\ref{fig:comp3} shows the density plot of the $\rm \Delta [Ce/Fe] (APOGEE - BAWLAS)$ for 94,522 stars in common. 
This metallicity-correlated offset seems to primarily be a result of using different \teff and \logg, calibrated for ASPCAP and uncalibrated for BAWLAS. Nevertheless, our sample of stars do not cover these metallicities, so we can overlook this trend.

\begin{figure}[ht]
\centering
\includegraphics[scale=0.45]{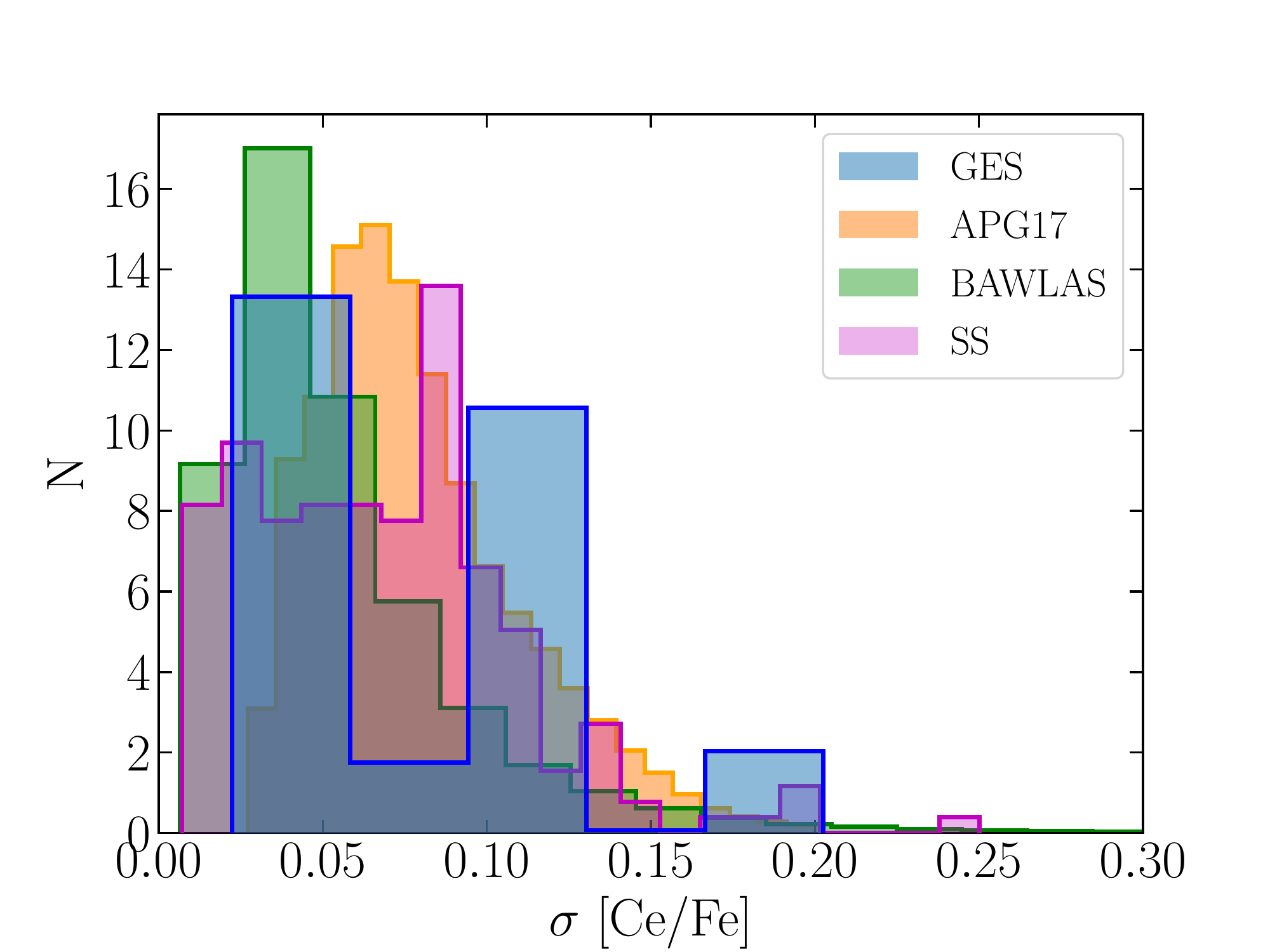}
\caption{[Ce/Fe] uncertainties distributions in APOGEE DR17, \citet{salessilva22}, \citet{hayes2022} and Gaia-ESO DR5. \label{fig:cefe_unc}}
\end{figure}

In all these cases, the measurements from APOGEE DR17 show a systematic offset of $\sim 0.1$~dex with a larger scatter at lower metallicity. This offset may be due the different pipelines, spectral range, and linelists used in the different surveys. 
Moreover, the uncertainties on the Ce abundance present in the APOGEE DR17 survey are larger than the uncertainties from the Gaia-ESO survey, \citet{salessilva22} and \citet{hayes2022} analysis (see their distributions in Fig.~\ref{fig:cefe_unc}). 
Despite of a large standard deviation of the Ce abundance differences among the samples, we can conclude that the measurements from the APOGEE DR17 show a rather good agreement with the Gaia-ESO, \citet{salessilva22} and \citet{hayes2022} studies. 

\subsection{Open clusters}
\label{sec:ocs}

\begin{figure*}[ht]
\centering
\includegraphics[scale=0.4]{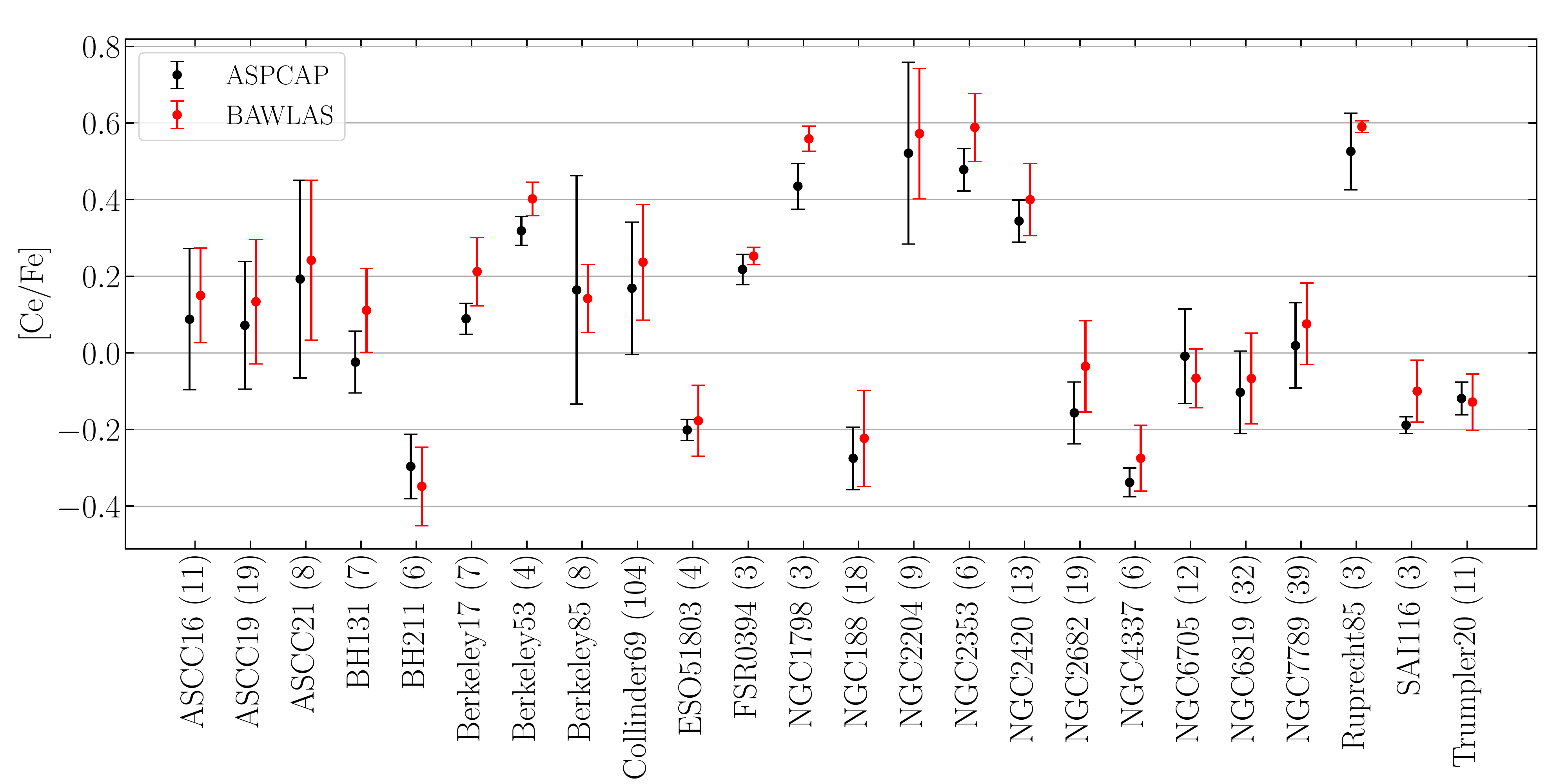}
\caption{The mean and standard deviation of [Ce/Fe] for the open clusters in common between ASPCAP (black) and BAWLAS (red) pipelines. The [Ce/Fe] mean in each cluster is represented with a dot and the standard deviation with the error bar. The number near to the cluster names represents the number of stars in each cluster. \label{fig:ocs}}
\end{figure*}

To understand if the use of the BAWLAS pipeline can provide an advantage to our work with respect to ASPCAP, we compute the mean and standard deviation of [Ce/Fe] for the member stars of open clusters in common between the two pipelines. 
The list of the open clusters and the catalogue of their members are published in \citet{myers22}. 
The values are shown in Fig.~\ref{fig:ocs}. We can see that the average values  of [Ce/Fe] computed with BAWLAS results are larger than those computed with ASPCAP, except in a few cases. This discrepancy is related to the offset that we discuss in the previous section.
In both pipelines, the standard deviation is large ($\sigma \sim 0.1$~dex) for almost all clusters, showing a spread in the Ce measurements among the member stars of the same cluster, which is not expected since they should be homogeneous. Such large standard deviations indicate that the Ce uncertainties are likely underestimated by both pipelines. 
Moreover, our results are similar using one or the other pipeline. Therefore, we choose to use the ASPCAP results for the following section.

\section{The [Ce/Fe] abundance ratio trends}
\label{sec:cefe}
In this section, we show the {\it Kepler}, TESS and K2 stars selected following the criteria presented in Sect.~\ref{quality_check}. These samples contain 
stars with a distribution in age and metallicity spanning [0, 14] Gyr and [$-$1,+0.4] dex, respectively. 
In addition, we remove outliers using a Huber regression \citep{huber64}, with an hyperparameter set to 3.

In the next subsections, we present two different planes to present the samples: [Ce/Fe] vs. [$\alpha$/M] and [Ce/Fe] vs. [Fe/H].

\subsection{[Ce/Fe] vs. [$\alpha$/M]}
\begin{figure}[ht]
\centering
\includegraphics[scale=0.35]{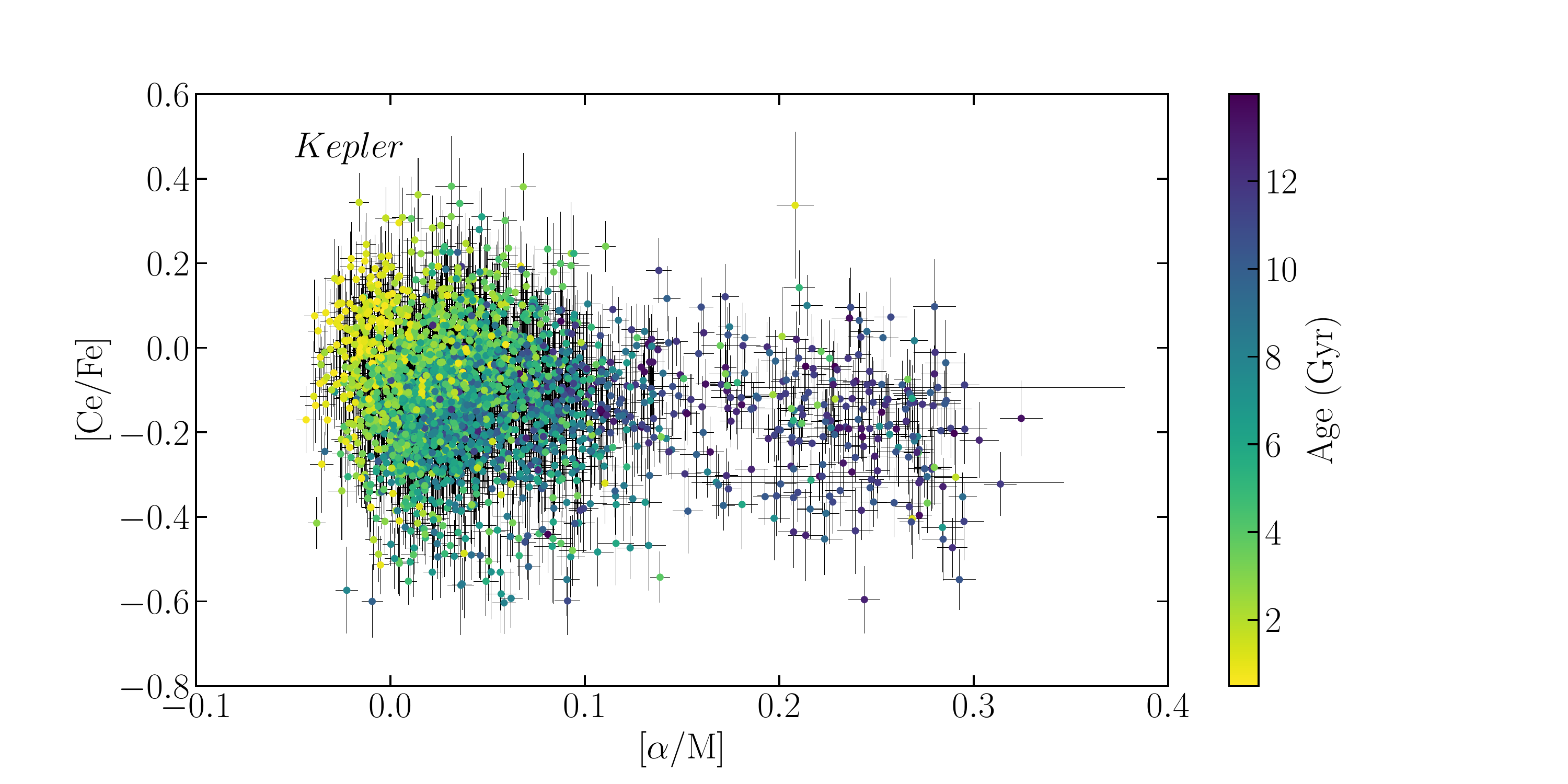}
\includegraphics[scale=0.35]{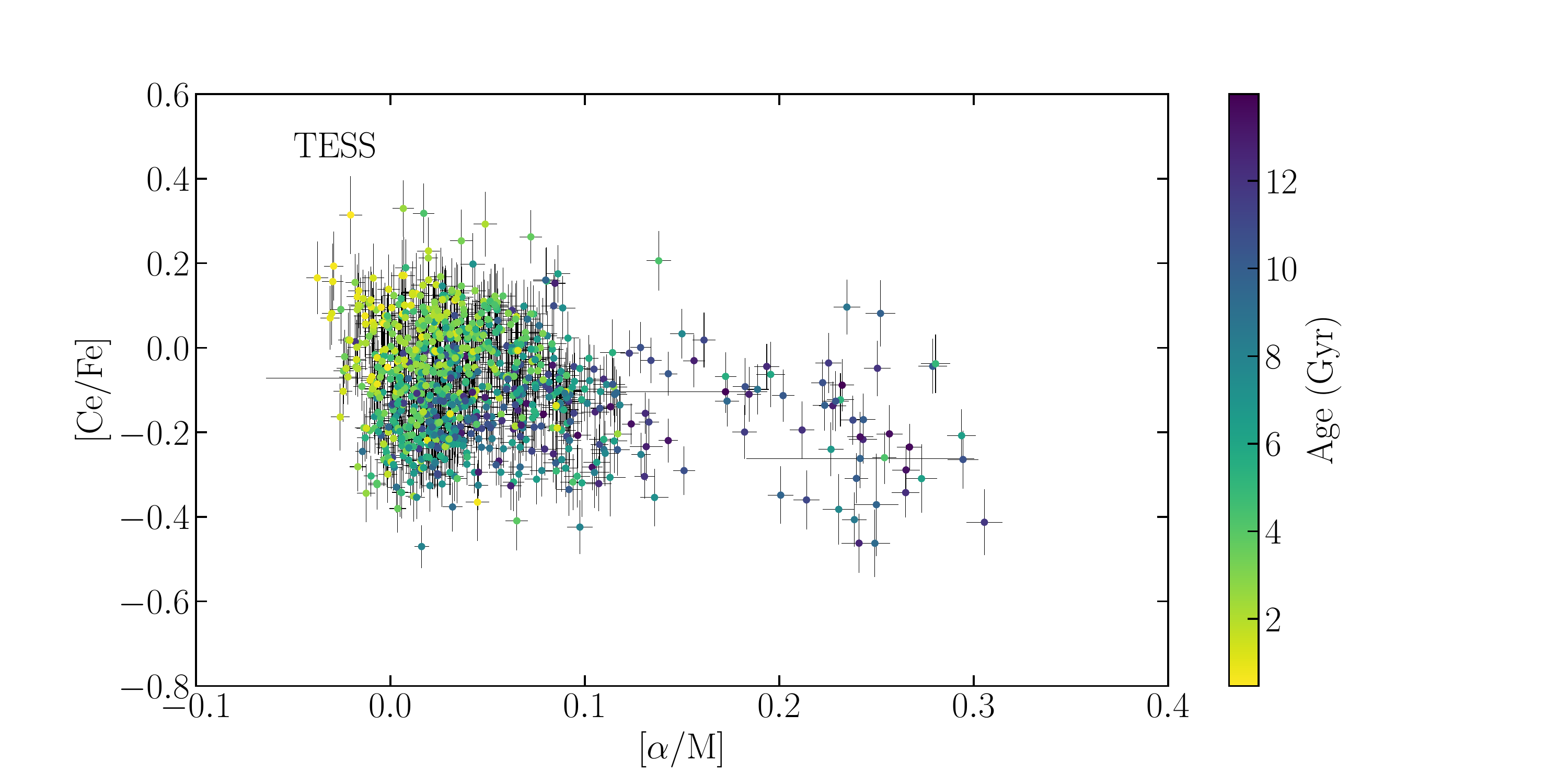}
\includegraphics[scale=0.35]{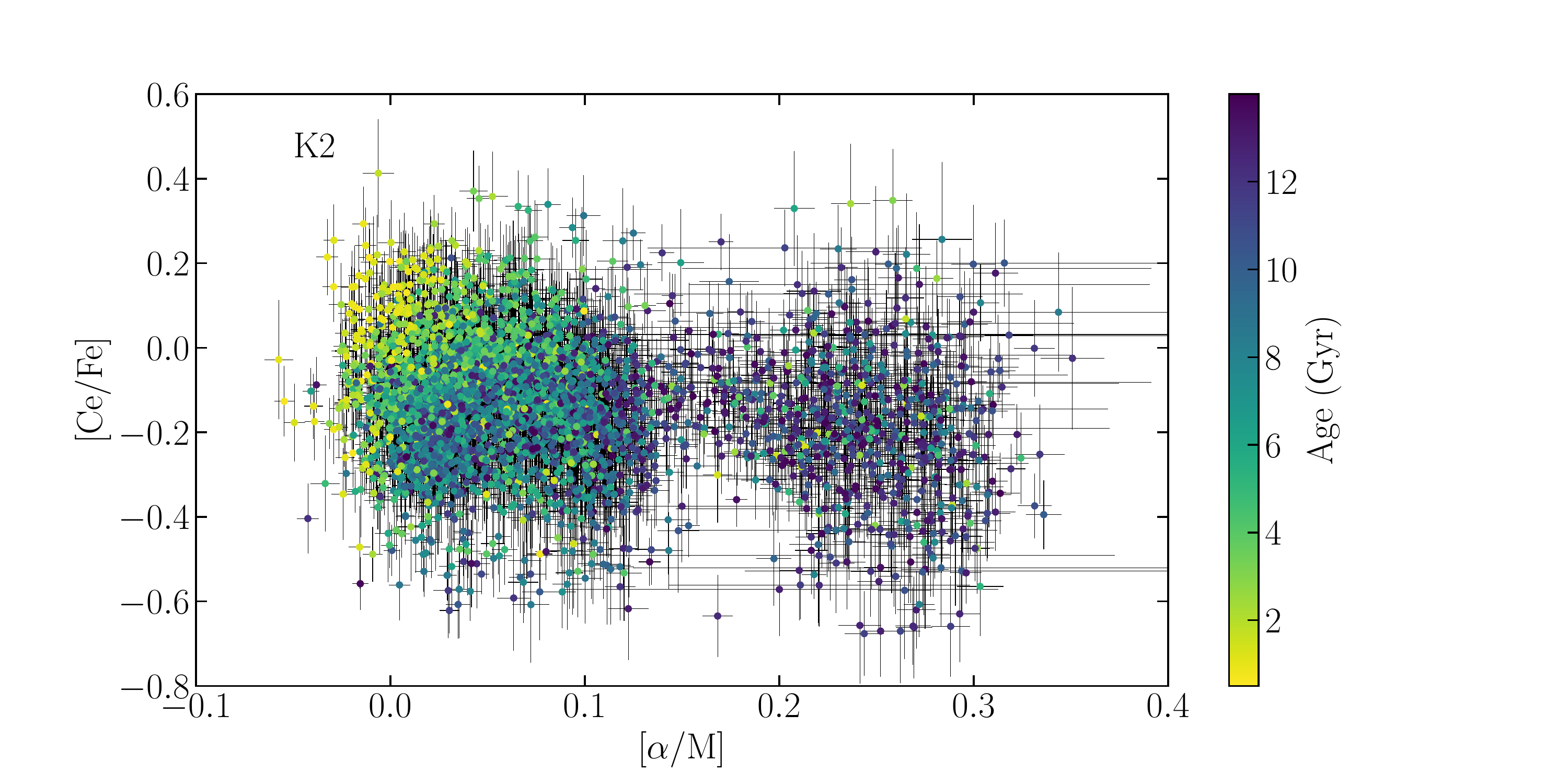}
\caption{[Ce/Fe] vs. [$\alpha$/M]. The data are respectively: {\it Kepler} (top panel), TESS (middle panel) and K2 (bottom panel) stars. The stars are colour-coded by stellar age. \label{fig:cefe_afe}}
\end{figure}

Figure~\ref{fig:cefe_afe} shows the [Ce/Fe]-[$\alpha$/M] for our three dataset, colour-coded by age. 
We use ASPCAP's [$\alpha$/M] as a proxy for [$\alpha$/Fe]. In our dataset, the difference between [$\alpha$/M] and an average [$\alpha$/Fe] defined using O, Mg, Si, S, and Ca over Fe is negligible (with a mean offset equal to $-0.002$ dex and a standard deviation of 0.016 dex).
In all datasets, we have two samples of stars, that display  two stellar populations: the low-$\alpha$ ([$\alpha$/M] $\lesssim$ 0.15 dex) and high-$\alpha$ sequence ([$\alpha$/M] $\gtrsim$ 0.15 dex). The number of high-$\alpha$ stars is larger for the K2 sample and lower for the TESS one. Additionally, the high-$\alpha$ sequence contains also the oldest stars in all three samples with a mean age $\sim 10-11$~Gyr \citep[see, e.g.,][for the age of the high-$\alpha$ population]{miglio21}. The other sequence, instead, spans a large range in the stellar age with also a gradient in age with increasing [Ce/Fe] and [$\alpha$/M]. Finally,  the high-$\alpha$ sequence has an average of [Ce/Fe] around $-0.2$ dex, while the low-$\alpha$ sequence has an average of [Ce/Fe] around $-0.1$ dex.

In the next sections, we focus on the low-$\alpha$ sequence since low and intermediate-mass AGB stars polluted mainly this stellar population. 
Moreover, it is more numerous with respect to the high-$\alpha$ sequence and allows us to compare our field stars with open clusters present in the literature \citep[e.g.,][]{salessilva22,viscasillas22,casamiquela21}. 
Furthermore, we do not make a selection in $z$, the height on the Galactic plane, because this cut would remove the old stars from our samples.

Here, we adopt a separation in [$\alpha$/M], using a piece-wise function to divide the populations. The function is an adjustment of the separation curve proposed by \citet{adibekyan12} for our data:

\begin{equation}
{\rm [\alpha/M]} = \begin{cases} 0.15 & {\rm [Fe/H]} \leq -0.3 \\ 
-0.34 \cdot {\rm [Fe/H]} + 0.05 & -0.3 < {\rm [Fe/H]} \leq 0 \\
0.05 & {\rm [Fe/H]} > 0 \end{cases}
\end{equation}

We choose this division following the most-adopted functions to separate low- and high-$\alpha$ sequences present in literature \citep[see e.g.,][]{adibekyan12,mikolaitis14}.

\subsection{[Ce/Fe] vs. [Fe/H] in the low-$\alpha$ sequence}
\begin{figure}[ht]
\centering
\includegraphics[scale=0.35]{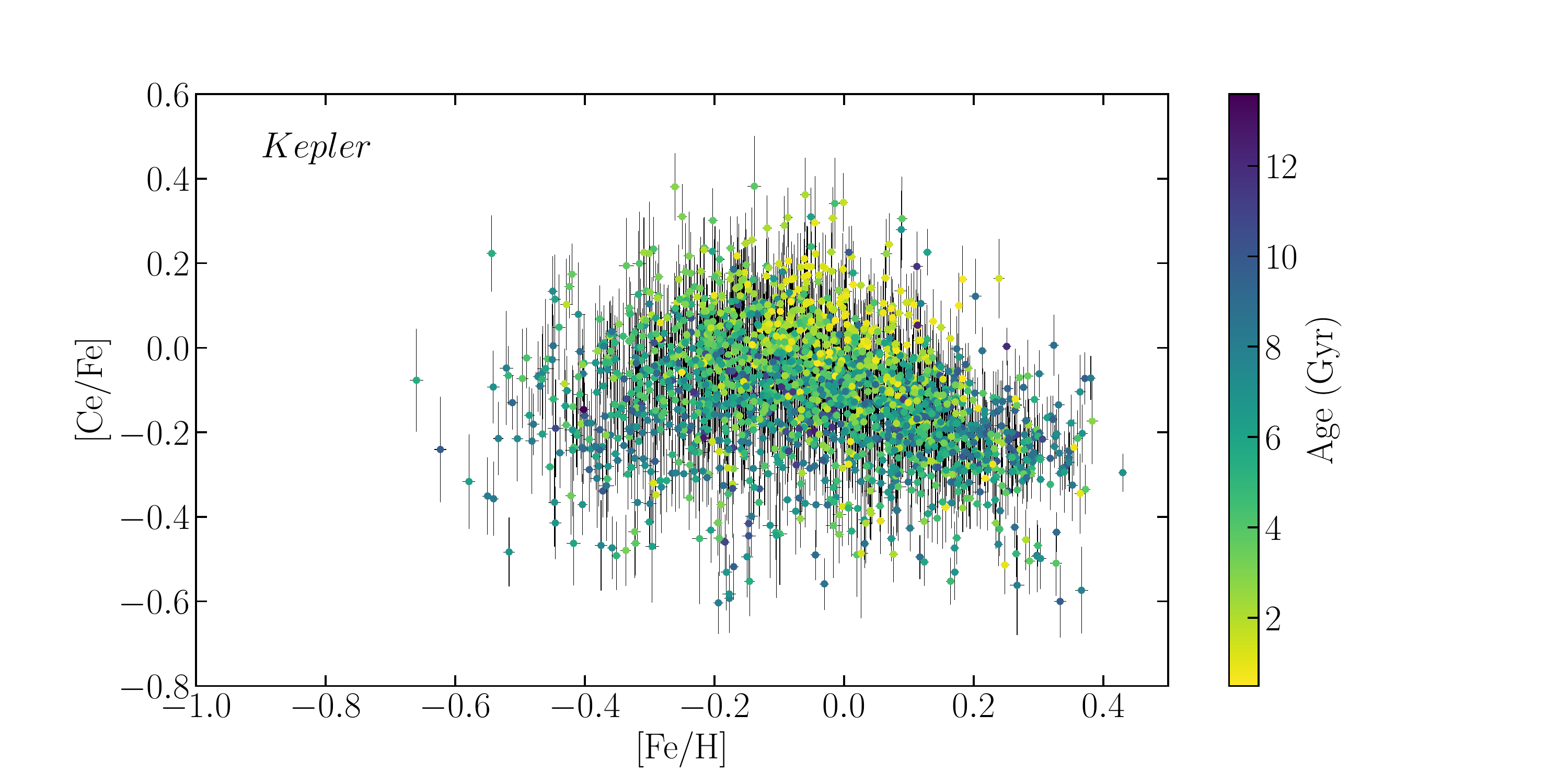}
\includegraphics[scale=0.35]{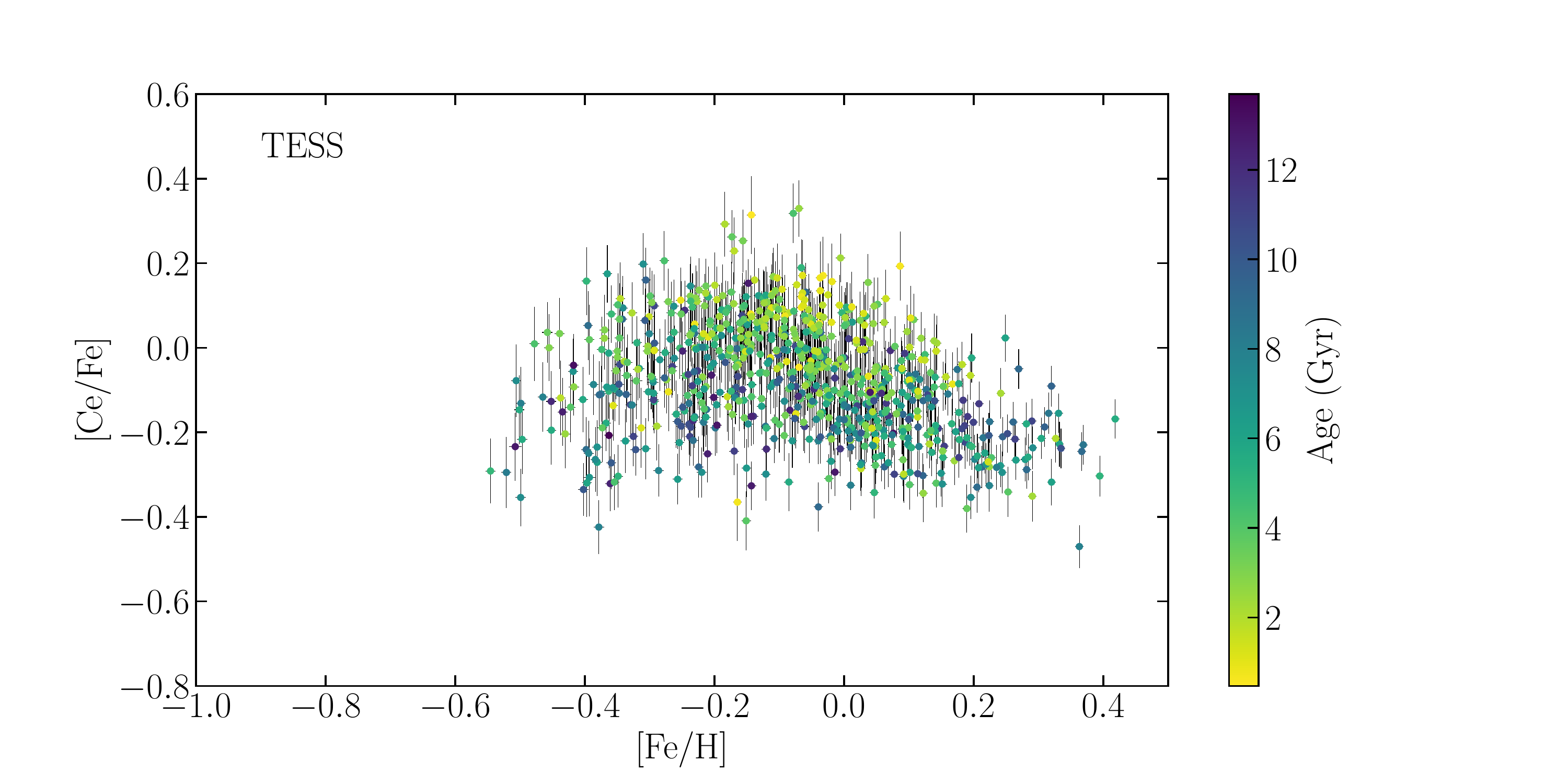}
\includegraphics[scale=0.35]{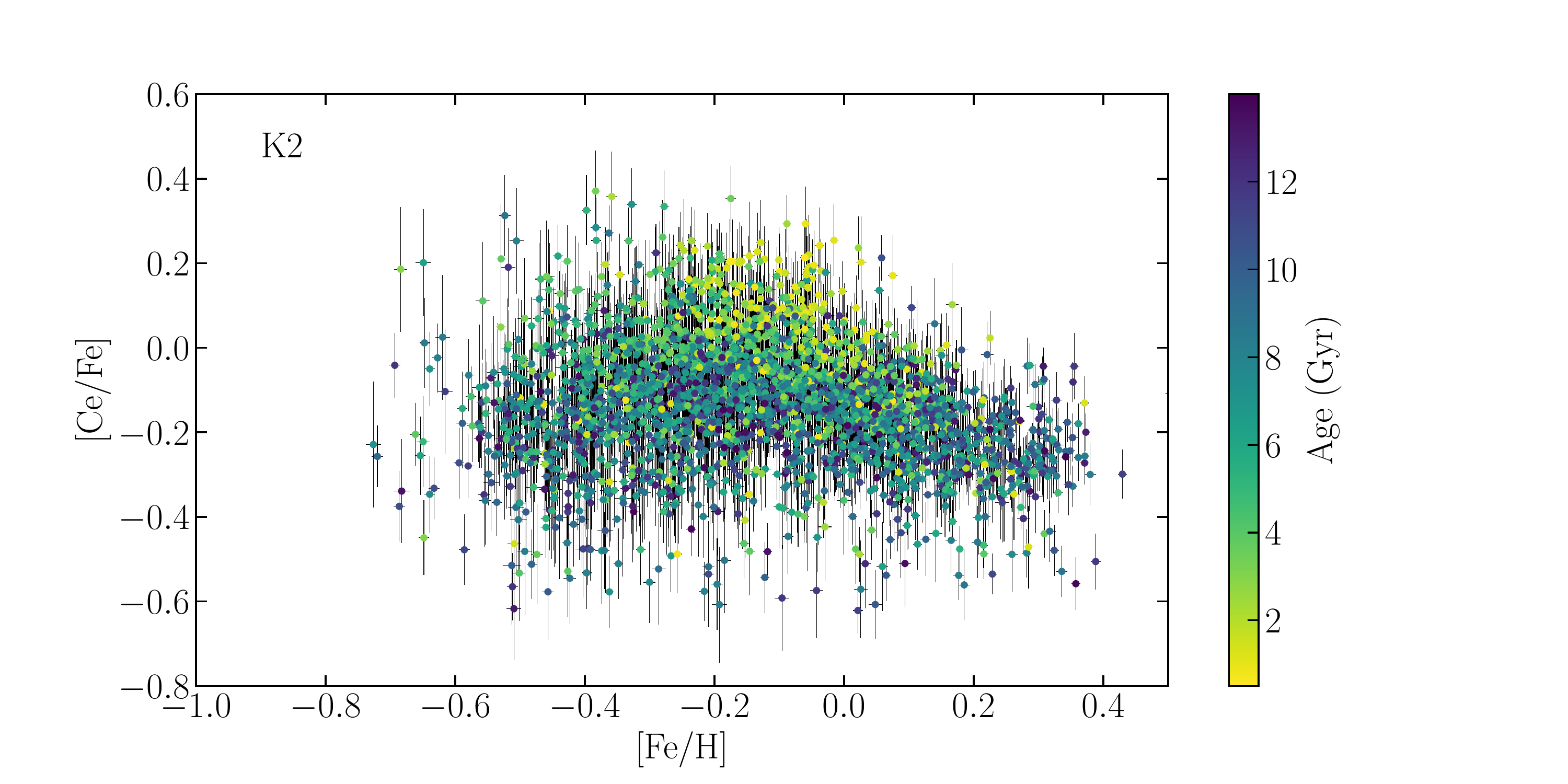}
\caption{[Ce/Fe] vs. [Fe/H] in low-$\alpha$ sequence. The data are respectively: {\it Kepler} (top panel), TESS (middle panel) and K2 (bottom panel) stars. They are colour-coded by stellar age. \label{fig:cefe_feh}}
\end{figure}

In Fig.~\ref{fig:cefe_feh}, we show the [Ce/Fe] trend in the low-$\alpha$ sequence as a function of metallicity for the three samples. 
They show the same shape:  an increasing [Ce/Fe] at increasing [Fe/H] with a maximum at $\sim -0.2$ dex in [Fe/H], and a drop in [Ce/Fe] at higher metallicities. The stars are also colour-coded by stellar age. At a given metallicity, the younger field stars show a [Ce/Fe] content larger than the older ones. 
A better visualisation of this trend is shown in the histograms in Fig.~\ref{fig:hist}. These histograms display the [Ce/Fe] binned in different ranges of stellar ages. The peaks of the histograms move towards lower [Ce/Fe] with increasing age. 

A similar behaviour is also shown in the samples of open clusters present in \citet{viscasillas22} and \citet{salessilva22}. In the former, they analysed 62 open clusters observed by the Gaia-ESO DR5 survey, while, in the latter, they study 42 open clusters observed by the APOGEE DR16 survey (with spectral analysis using BACCHUS).

As described in the introduction, the predominant fraction of s-process elements like Ce is produced by long-lived stars \citep[$1.5-3.0 \, M_{\odot}$, see, e.g.,][]{cristallo09,cristallo11}. 
This fact, together with the secondary nature (dependence on metallicity) of the s-process elements, determines the behaviour observed in the figure. Chemical evolution models present in literature (\cite{pranzos2018,grisoni2020}, the three-infall model in \cite{contursi22}) show a banana shape with a rise in [Ce/Fe], a peak and then a decline, similar to that seen in Fig.~\ref{fig:cefe_feh}.

In addition, the s-process production of Ce in AGB stars is 
strongly dependent on the metallicity \citep{busso01,vescovi20}. It depends on (i) the number of iron nuclei as seeds for the neutron captures, and (ii) the flux of neutrons. The former decreases with decreasing metallicity, while the latter increases because it depends (approximately) on $\rm ^{13}C$/$\rm ^{56}Fe$, which increases with decreasing metallicity ($\rm ^{13}C$ is a primary element, and does not depend on metallicity). This means there are more neutrons per seed in low-metallicity AGB stars and less in high-metallicity AGB stars.
Therefore, stellar evolution models predict lower [Ce/Fe] in higher-metallicity AGB stars \citep{cristallo15,karakas2016,battino19}, implying the trend that we see in Fig.~\ref{fig:cefe_feh}.


\section{The time evolution of Ce abundance in the low-$\alpha$ sequence}
\label{sec:tempevol}

In this section, we study the temporal evolution on the Ce abundance in the low-$\alpha$ sequence. We explore the [Ce/Fe] and [Ce/$\alpha$] trends as a function of age and position in the Galactic disc.

\subsection{[Ce/Fe] vs. Age}
\label{sec:cefe_age}

\begin{figure}[ht]
\centering
\includegraphics[scale=0.32]{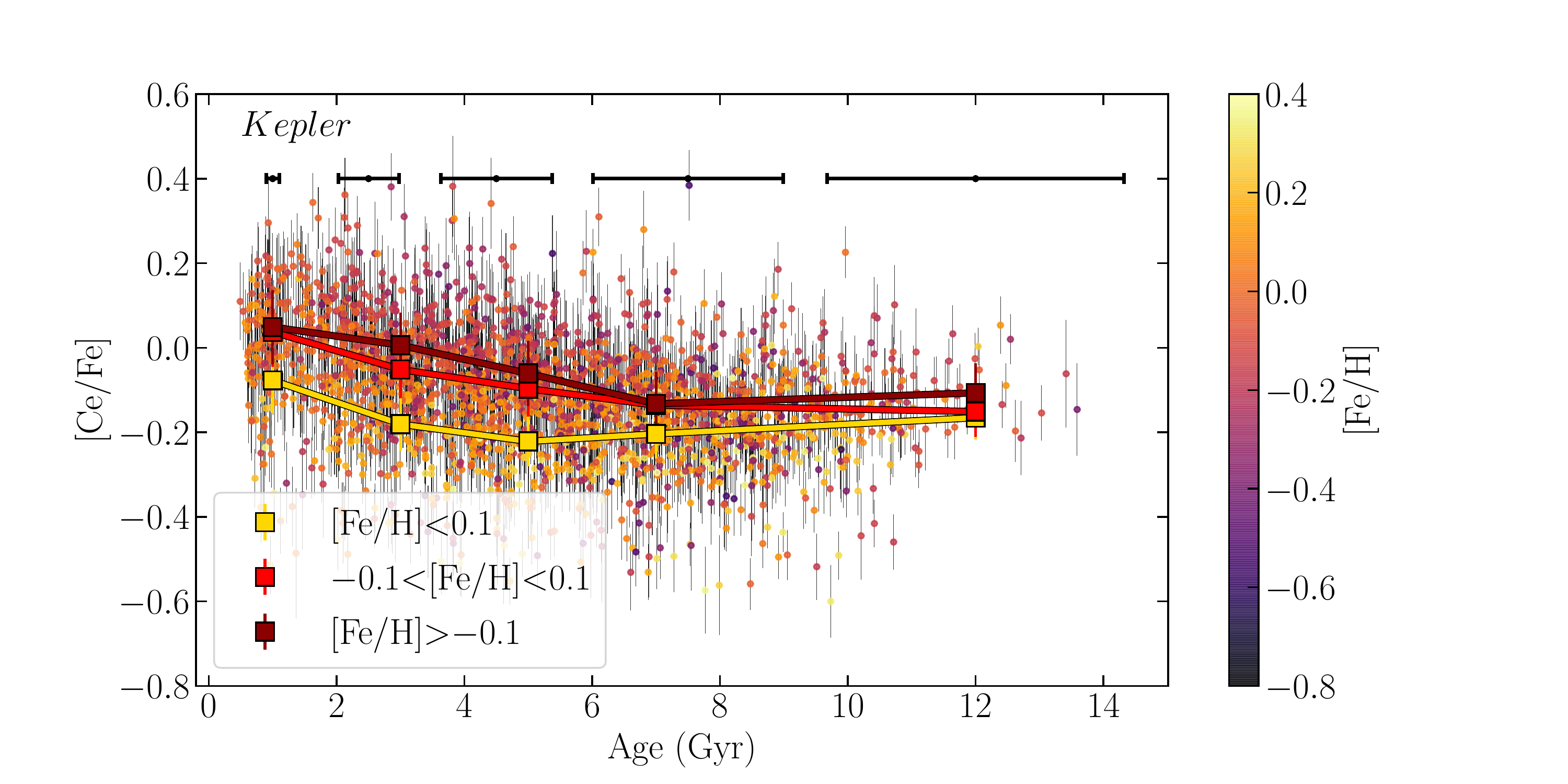}
\includegraphics[scale=0.32]{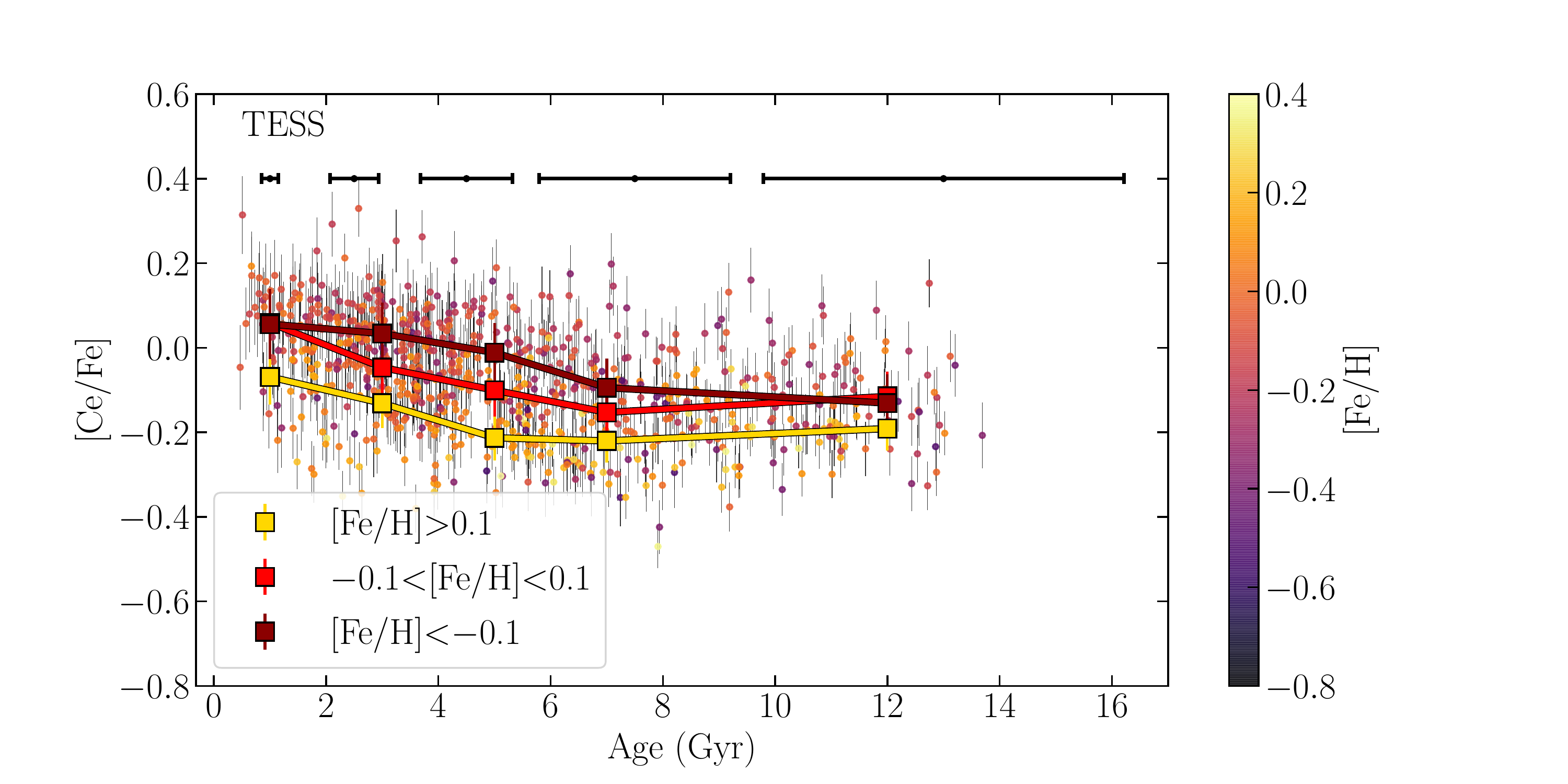}
\includegraphics[scale=0.32]{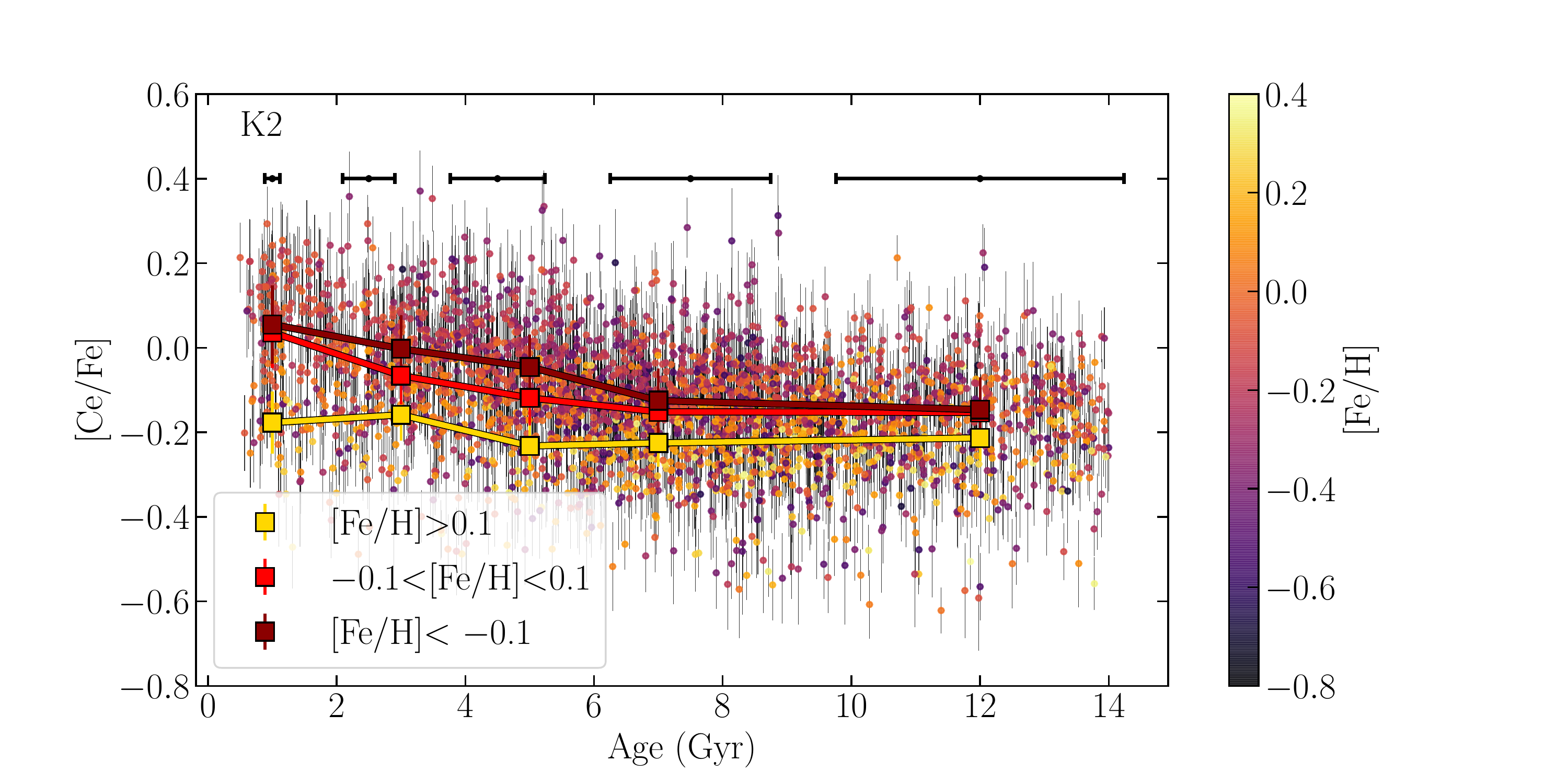}
\caption{[Ce/Fe] vs. stellar age for the three samples. The data are colour-coded by metallicity. The three lines represent the [Ce/Fe] mean in different bins of metallicity and age. The error badget of stellar age is shown in the upper part of the panels. \label{fig:cefe_age_all}}
\end{figure}

Figure~\ref{fig:cefe_age_all} shows the  correlation of [Ce/Fe] with the stellar age for our three samples, colour-coded by metallicity. The three lines in the figure represent the mean [Ce/Fe] in three different bins of [Fe/H]: [Fe/H] $<$ $-$0.1, $-$0.1 $\leq$ [Fe/H] $\leq$ 0.1, [Fe/H] $>$ 0.1 dex. 
These three lines clearly illustrate the metallicity dependence of the [Ce/Fe]-age relation. In particular, at a given age, metal-poor stars have a higher [Ce/Fe] ratio than the metal-rich ones. 
Moreover, [Ce/Fe] increases with decreasing ages and becomes almost flat for ages older than 6 Gyr. The same behaviour is shown in \citet{salessilva22} for the open clusters. 
However, in this figure and in the work of \citet{salessilva22}, the location of the stars in the Galactic disc is not taken into account. 

The majority of our  stars are located in the solar neighbourhood (7.5 $<$ R$_{GC}$ $<$ 8.5 kpc). Nonetheless, they can come from the inner or outer regions of the Galactic disc and transit in the solar vicinity, or can be migrated in previous epochs due to the change of their eccentricity through radial heating, or their angular momentum. To better understand their provenance, we can compute the so-called guiding radius, $R_{g}$, that is the radius of a circular orbit with specific angular momentum $L_{z}$.
Moreover, the adoption of $R_{g}$, instead of the Galactocentric radius $R_{GC}$, can mitigate the blurring effect due to epicyclic oscillations around the guiding radius \citep{schonrich09}. However, it cannot overcome the migrating effect due to the churning, which can change $R_{g}$ due to interactions with spiral arms or bars \citep{sellwood02,binney08}. 

\begin{figure}[ht]
\centering
\includegraphics[scale=0.45]{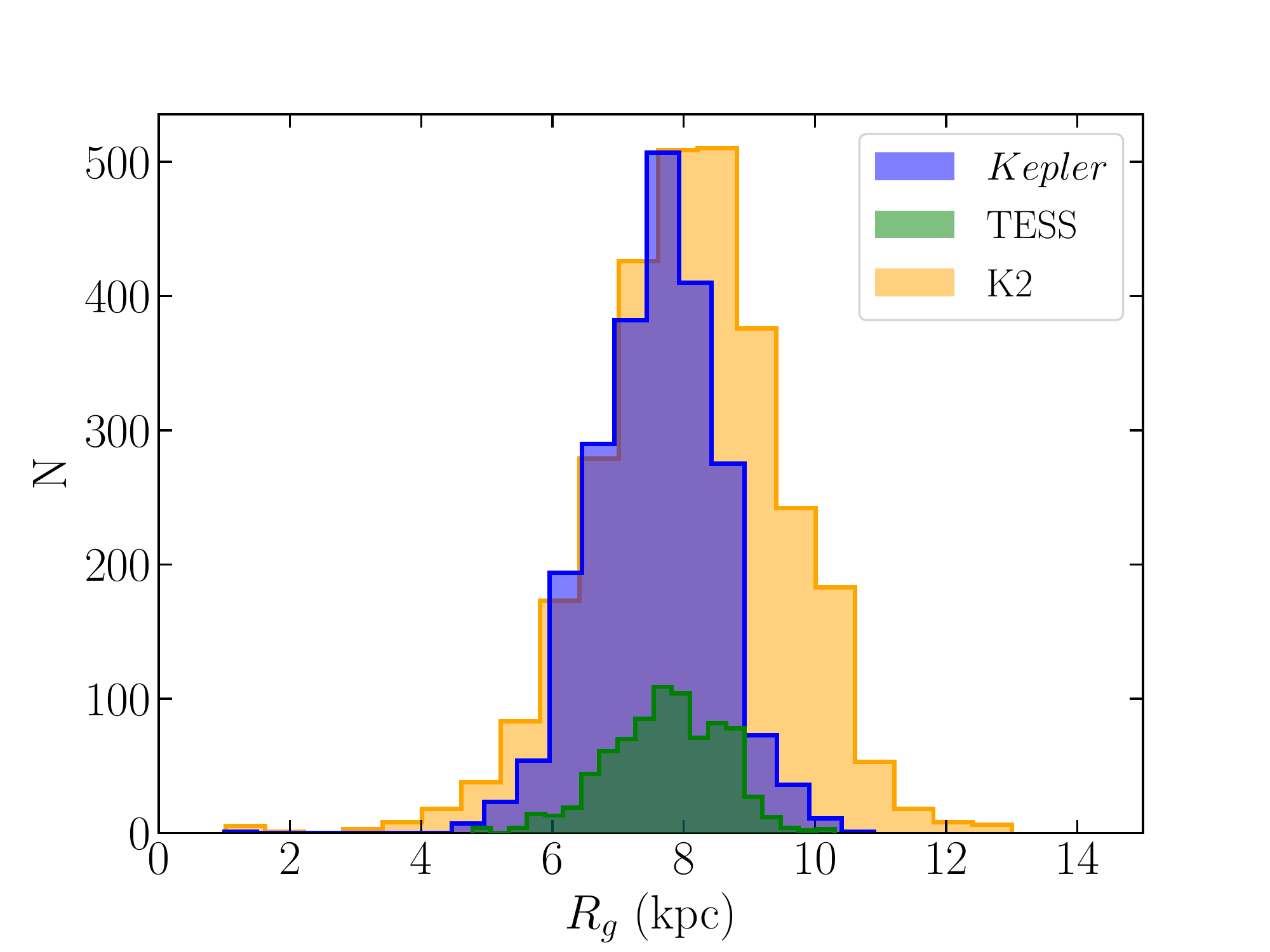}
\caption{Distribution of the $R_{g}$ of the three selected samples. \label{fig:histRg}}
\end{figure}

$R_{g}$ is computed from the stellar orbits obtained  using the \texttt{GalPy} package of Python, in which the model \texttt{MWpotential2014} for the gravitational potential of the Milky Way is assumed \citep{bovy15}.
Through the astrometric information by Gaia EDR3, distances from \citet{bailerjones21}, an assumed solar Galactocentric distance $R_{0} = 8$ kpc, a height above the plane $z_{0} = 0.025$ kpc \citep{juric08}, a circular velocity at the solar Galactocentric distance equal to $V_{c} = 220 \rm ~km~s^{-1}$ and the Sun's motion with respect to the local standard of rest $[U_{\odot}, V_{\odot}, W_{\odot}] = [11.1, 12.24, 7.25] \rm ~km~s^{-1}$ \citep{schonrich10}, we obtain the orbital parameters, among which the guiding radius $R_{g}$. 
The distribution of $R_{g}$ of the three dataset is shown in Fig.~\ref{fig:histRg}. 
The K2 sample shows a larger distribution with respect to those of {\it Kepler} and TESS: those of the latter are located between $5 \lesssim R_{g} \lesssim 10$~kpc, while K2 reaches $R_{g} \sim 4$~kpc in the inner disc and  $R_{g} \sim 12$~kpc in the outer disc, as expected from the different field locations and distances reached.


\begin{figure*}[ht]
\centering
\includegraphics[scale=0.55]{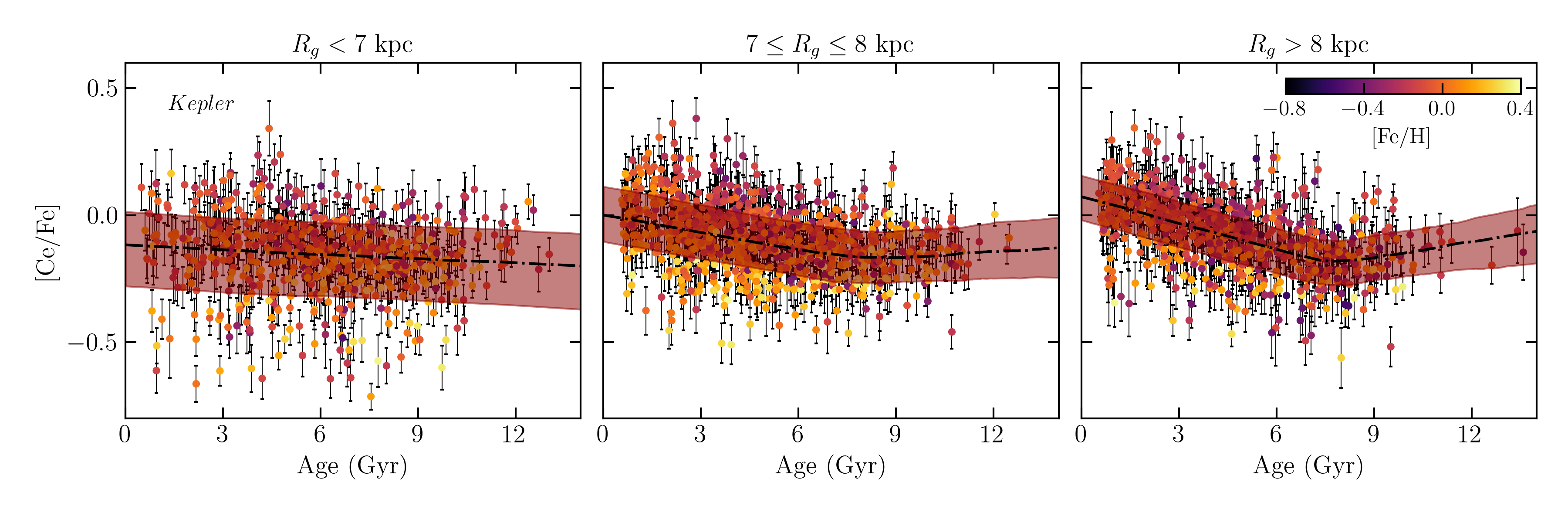}
\includegraphics[scale=0.55]{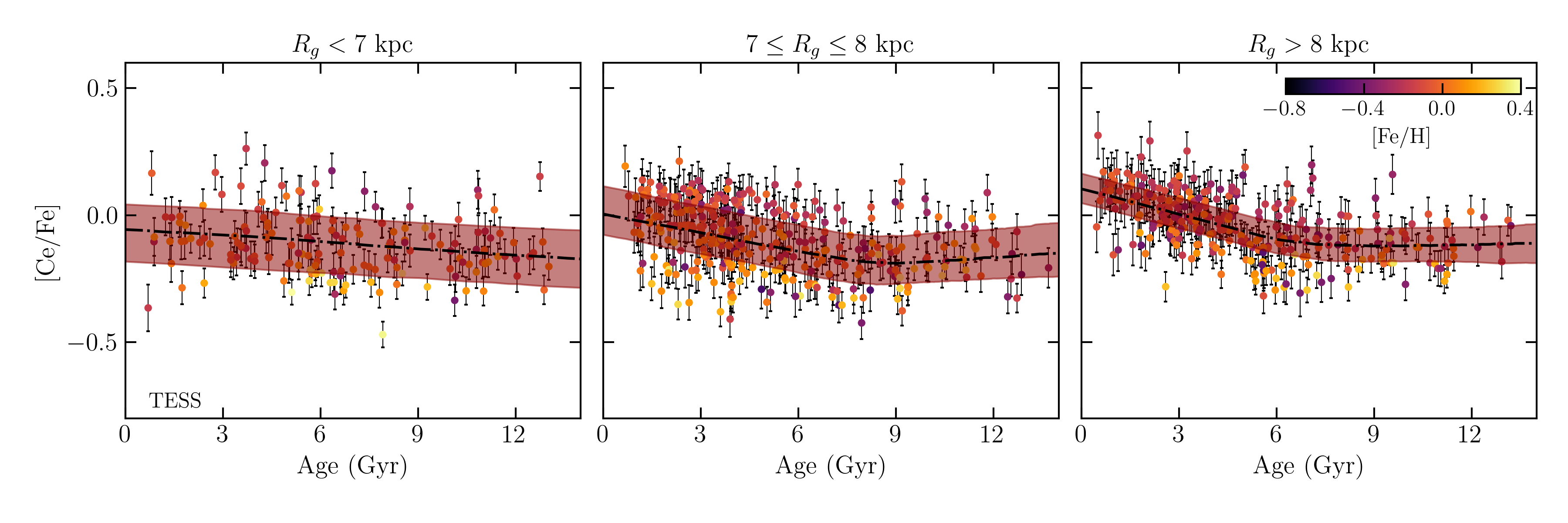}
\hspace*{-1.5cm}\includegraphics[scale=0.5]{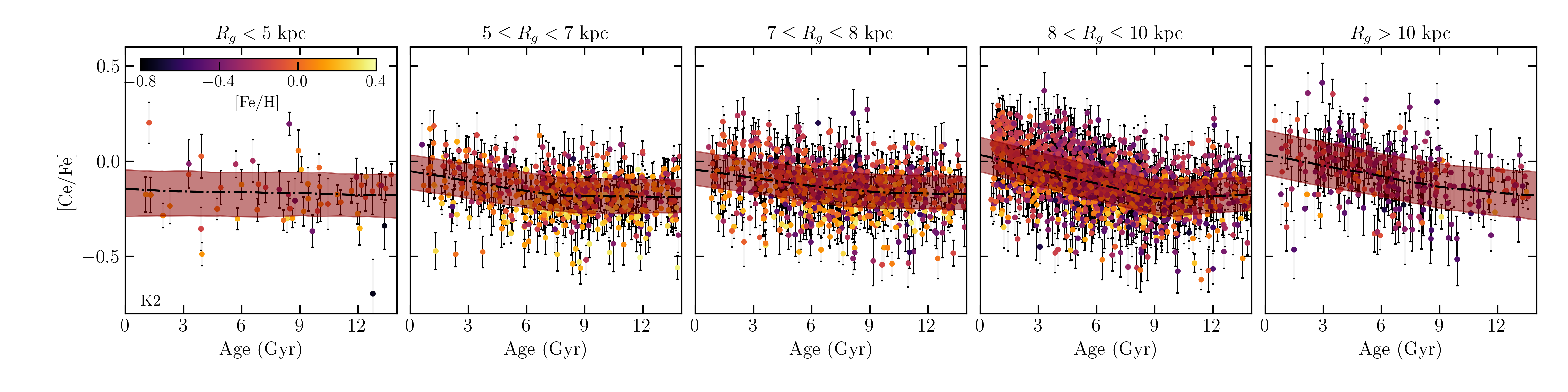}
\caption{[Ce/Fe] vs. stellar age in different bins of R$_{g}$. The {\it Kepler} sample is shown in the upper panel, the TESS sample in the middle and the K2 one in the bottom one. The data are colour-coded by metallicity. The dash-dotted line represents the best fit and the shaded area the 68\% confidence interval plus intrinsic scatter. \label{fig:cefe_age}}
\end{figure*}

Figure~\ref{fig:cefe_age} shows the same plane of Fig.~\ref{fig:cefe_age_all} for the {\it Kepler}, TESS and K2 samples, divided in three different bins of guiding radius: $R_{g} < 7$ kpc, $7 \leq R_{g} \leq 8$ kpc, $R_{g} > 8$ kpc. Since K2 has a larger range in $R_{g}$ with respect to {\it Kepler} and TESS (see Fig.~\ref{fig:histRg}), we add two more bins for this dataset, not covered by the other two data samples: $R_{g} < 5$ kpc and $R_{g} > 10$ kpc.

In order to visualise in a more quantitative way the different trends in the different bins and the flattening at old age, we model the distribution of [Ce/Fe] vs. stellar age through a broken-line as follows:

\begin{equation}
{\rm [Ce/Fe]} = \begin{cases} m_{1} \cdot {\rm Age} + c, & Age \leq k \\ n_{1} \cdot (Age - k) + (m_{1} \cdot k + c), & Age > k \end{cases}
\end{equation}

Through a Markov Chain Monte Carlo (MCMC) procedure, we derive the posterior distributions of the parameters of the fitting $m_{1}$, $c$, $n_{1}$ and $k$, where $k$ is the switch point of the fitting line. In our model, we take into account the uncertainties on [Ce/Fe] and stellar age. We also include in our calculation an intrinsic scatter of the relation, $\epsilon$. 

In this procedure, we use uniform priors for $m_{1}$, $c$, $n_{1}$, $k$ and $\epsilon$, with limits for $k$ ($4 < k < 10$ Gyr) and $\epsilon$ ($\epsilon > 0$). We run the simulation with 10,000 samples, 1,000 of which are used for burn-in. The script is written in Python using the \texttt{emcee} package \citep{emcee}.

In order to decide if the broken-line was the best model for our data, we compare a broken-line and a single-line given the data by computing the Bayesian Information Criterion (BIC). The model with the lowest BIC is generally the most reasonable one. In the bins with $R_{g}$ larger than 7 kpc, the broken-line provides a lower BIC, while in the bins of $R_{g} < 7$~kpc for the {\it Kepler} and TESS samples and $R_{g} < 5$~kpc for the K2 one, the single-line provides the lowest BIC. For the range $5 \leq R_{g} < 7$~kpc of the K2 sample, the broken-line provides a marginally lower BIC than the single-line. 
We adopt the model with the lowest BIC respectively in each region. 

The convergence of the Bayesian inference is checked against the traces of each parameter and their auto-correlation plots. The spread (68\% confidence interval plus intrinsic scatter) of the models resulting from the posteriors are represented in Fig.~\ref{fig:cefe_age} with the red shaded area, while their values are reported in Table~\ref{tab:mcmc_cefe}. 

Looking at the [Ce/Fe] trends in the three samples, we notice a different behaviour, in particular in the young regime, between inner and outer regions. The latter usually have higher [Ce/Fe] than the former at a given age (the slope is more negative, see the second column on Table~\ref{tab:mcmc_cefe}). Moving towards the outer regions the [Ce/Fe]-age trends become steeper and steeper in all our datasets. Dividing the samples by guiding radius makes clear that there is a strong dependence of the [Ce/Fe] abundances on the location of the stars. The behaviour of [Ce/Fe] with $R_{g}$ seems to be complex and likely related to different star formation histories and to metallicity dependency of the stellar yields.

In more detail, the left panels of Fig.~\ref{fig:cefe_age} ($R_{g}<7$ kpc) show a large fraction of {\it Kepler} and K2 stars coming from the inner disc with respect to the TESS stars. Moreover, the trends of [Ce/Fe] with stellar age in these bins of $R_{g}$ are quite flat with a slope of $\sim -0.01$~dex/Gyr, against the slopes in the outer regions of $\sim -0.02/-0.03$~dex/Gyr.
These trends could be due to the more intense star formation efficiency in the inner regions of the Galactic disc, where high values of [Ce/Fe] are reached at earlier epochs. 

In the other panels with $R_{g}>7$ kpc, there is a clear increasing trend with decreasing age. The slope in the young regime becomes steeper towards the outer regions ($R_{g}>8$ kpc) in all datasets. 
In the solar vicinity and outer disc, where the star formation rate is less intense with respect to the inner regions, the increasing trends with decreasing age can be due to the recent contribution of low- and intermediate-mass AGB stars to the Galactic enrichment. Instead, the flattening of the slope at ages older than 6-8 Gyr (see the parameter $k$ in Table~\ref{tab:mcmc_cefe}) might reflect the lower contribution of low- and intermediate-mass AGB stars in early epoch of Galaxy formation, because they have yet to start to contribute significantly.
We also test if the flattening of the data in the oldest regime could be due to their larger uncertainties on age. We generate mock data with a distribution similar to ours, an uncertainty of 30\% on age and an intrinsic scatter of 0.05~dex, using a single-line model (see Appendix~\ref{app:mock}). Applying the same MCMC procedure as for the two models (broken-line and single-line), we obtain a lower BIC for the single-line. This suggests that the flattening is not a consequence of the larger uncertainties on the oldest age, but a consequence of the Galactic chemical evolution.

\begin{table*}[h]
 \caption{Parameters of the MCMC fitting for the [Ce/Fe]-age relations in the {\it Kepler}, TESS and K2 samples.}
\begin{center}
\begin{tabular}{cccccc}
\hline
Bin & $m_{1}$~(dex/Gyr) & $c$~(dex) & $n_{1}$~(dex/Gyr) & $k$~(Gyr) & $\epsilon$~(dex)  \\
\hline
& & & {\it Kepler} &  &\\
\hline
$R_{g} <7$ kpc            & ${-0.006} \pm {0.002}$ & ${-0.122} \pm {0.016}$ & -- & -- & ${0.139} \pm {0.005}$  \\
$7 \leq R_{g} \leq 8$ kpc & $-0.022 \pm 0.003$ & $ 0.007 \pm 0.011$ & $0.005^{+0.013}_{-0.009}$ &  $7.6^{+0.9}_{-0.9}$  &  $0.099 \pm 0.004$  \\
$R_{g}>8$ kpc             & $-0.035 \pm 0.003$ & $ 0.071 \pm 0.011$ & $0.018 \pm 0.012$         &  $7.6^{+0.4}_{-0.6}$  &  $0.086 \pm 0.005$  \\
\hline
 & & & TESS &  & \\
\hline
$R_{g} <7$ kpc              & ${-0.008} \pm {0.003}$ & ${-0.062} \pm {0.023}$ & -- & -- & ${0.104} \pm {0.008}$  \\
$7 \leq R_{g} \leq 8$ kpc & $-0.025 \pm 0.004$ & $ 0.016 \pm 0.016$ & $ 0.008^{+0.012}_{-0.010}$ &  $7.9^{+1.0}_{-0.8}$  &  $0.095 \pm 0.006$  \\
$R_{g}>8$ kpc             & $-0.034 \pm 0.004$ & $ 0.101 \pm 0.016$ & $0.0005^{+0.012}_{-0.008}$ &  $6.8^{+1.3}_{-0.9}$  &  $0.065 \pm 0.007$  \\
\hline 
 & & & K2 &  & \\
 \hline
$R_{g} < 5$~kpc  & ${-0.001} \pm {0.005}$ & ${-0.158} \pm {0.047}$ & $--$ &  $--$ & ${0.110} \pm {0.017}$ \\
$5 \leq R_{g} < 7$~kpc  & ${-0.015} \pm {0.004}$ & ${-0.058} \pm {0.019}$ & ${0.000} \pm {0.004}$& ${7.93}^{+0.93}_{-1.38}$ & ${0.088} \pm {0.005}$  \\
$7 \leq R_{g} \leq 8$~kpc  & ${-0.015} \pm {0.003}$ & ${-0.044} \pm {0.017}$ & ${-0.004} \pm {0.004}$& ${7.85}^{+1.51}_{-1.38}$ & ${0.099} \pm {0.004}$  \\
$8 < R_{g} \leq 10$~kpc    & ${-0.025} \pm {0.002}$ & ${ 0.043} \pm {0.009}$ & ${ 0.006} \pm {0.006}$& ${9.18}^{+0.49}_{-0.52}$ & ${0.097} \pm {0.003}$  \\
$R_{g} > 10$ kpc          & $-0.020 \pm 0.005$ & $ 0.050 \pm 0.003$ & $-0.009^{+0.012}_{-0.008}$  &  $8.14^{+1.4}_{-2.5}$  &  $0.115 \pm 0.008$  \\
\hline
\end{tabular}
\end{center}
\label{tab:mcmc_cefe}
\end{table*}

\subsection{[Ce/$\alpha$] as a chemical clock}

The ratios of s-process elements to $\alpha$-elements are widely studied as age indicators because they show a steeper trend with age than [s/Fe].


In this section, we investigate  [Ce/$\alpha$]-age trends for the low-$\alpha$ sequence stars in the {\it Kepler}, TESS, and K2 samples, in the same $R_{g}$ bins shown in Sect.~\ref{sec:cefe_age}. 
The $\alpha$ elements we take into account in this work are the same studied by \citet{salessilva22} for their open clusters: O, Mg, Si and Ca. In this way, we can compare the trends present in our low-$\alpha$ sequence stars with their open clusters.

\begin{figure*}[ht]
\centering
\includegraphics[scale=0.6]{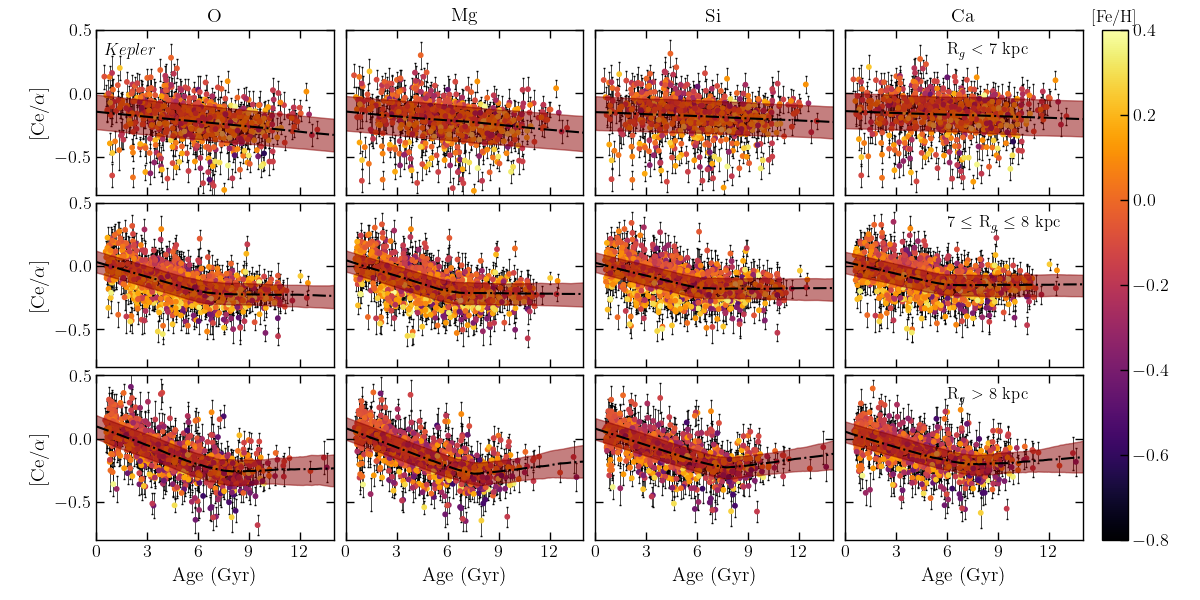}
\includegraphics[scale=0.6]{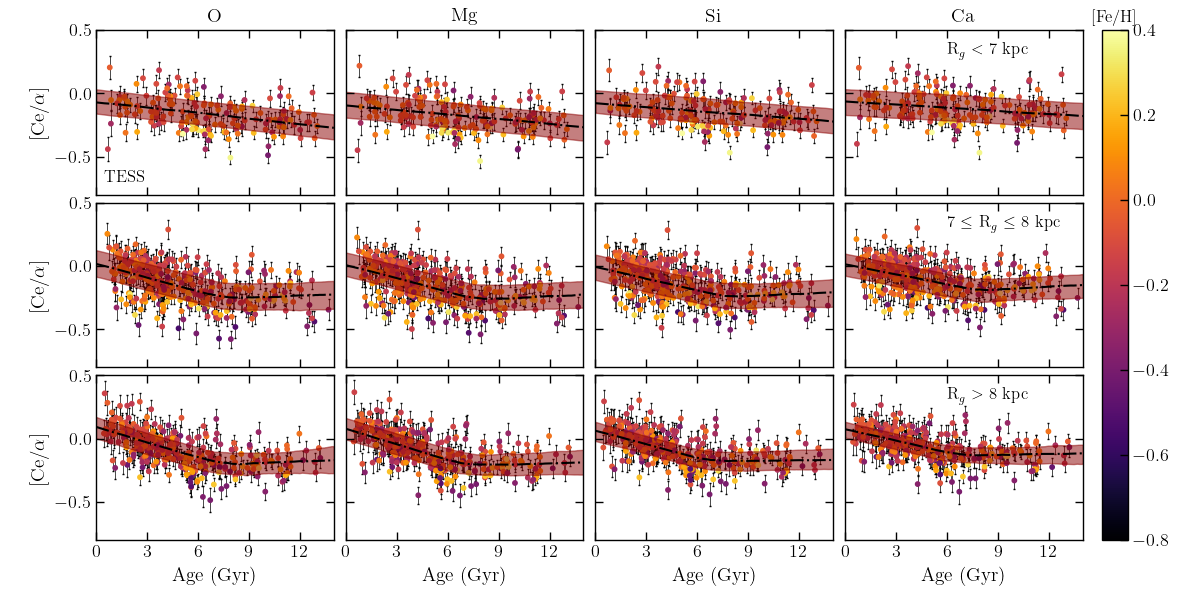}
\caption{[Ce/$\alpha$]-age planes for {\it Kepler} (top panels) and TESS (bottom panels). The points are colour-coded by [Fe/H]. The black dash dotted line represents the best fit and the red shaded area the 68\% confidence interval plus intrinsic scatter. \label{fig:cex_keptess}}
\end{figure*}

\begin{figure*}[ht]
\centering
\includegraphics[scale=0.6]{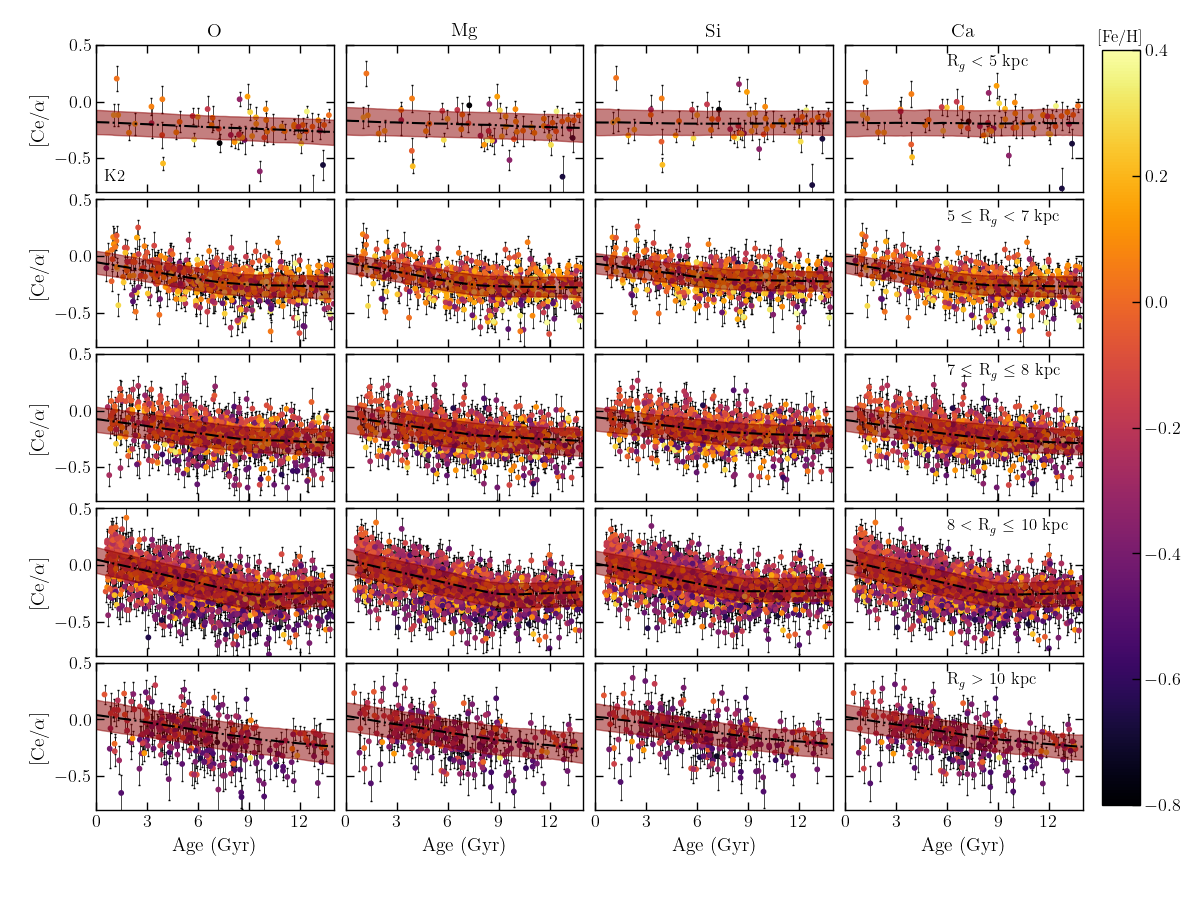}
\caption{[Ce/$\alpha$]-age planes for the K2 sample. The symbols are the same of Fig.~\ref{fig:cex_keptess}.  \label{fig:cex_K2}}
\end{figure*}

In order to study the [Ce/$\alpha$] trends in a more quantitative way, we model their distributions at different $R_{g}$, taking also into account the [Fe/H] dependence \citep[following the work by][]{delgado19,casali20,viscasillas22} as follows:

\begin{equation}
{\rm [Ce/\alpha]} = \begin{cases} m_{1} \cdot {\rm Age} + m_{2} \cdot {\rm [Fe/H]} + c, & Age \leq k \\ n_{1} \cdot (Age - k) + (m_{1} \cdot k + c + m_{2} \cdot {\rm [Fe/H]}), & Age > k \end{cases}
\end{equation}

In our calculation, we take also into account the uncertainties on [Ce/$\alpha$], [Fe/H] and stellar age. We adopt a single-line for the bin of $R_{g} < 7$~kpc for {\it Kepler} and TESS and the bin of $R_{g} < 5$~kpc for K2 where their BIC is lower.

The best fits with the spread (68\% confidence interval plus intrinsic scatter) of the models resulting from the posteriors are represented in Figs.~\ref{fig:cex_keptess} and \ref{fig:cex_K2} with a black line and a red shaded area. Their values are shown in Table~\ref{tab:mcmc}.

The [Fe/H] colour-coding of these stars show a dependence on metallicity. We can see that stars with lower metallicity have higher [Ce/$\alpha$] at a given age. Furthermore, the stars at higher metallicity show an almost flat trend with respect to the metal poor ones (see Fig.~\ref{fig:CeX_fehbins} for a better visualisation of these trends). The same behaviour is displayed by \citet{salessilva22} and \citet{viscasillas22} for their samples of open clusters. 
 
This result was already shown in \citet{feltzing17}, \citet{delgado19}, \citet{casali20}, \citet{magrini21} for [Y/$\alpha$]. \citet{casali20}, \citet{magrini21} and \citet{viscasillas22} suggested that the metallicity dependence is due to the different star formation history and the non-monotonic dependence of s-process yields on [Fe/H]. In particular, \citet{magrini21} investigated the different metallicity dependence of s-process AGB yields if we include or not the magnetic-buoyancy-induced mixing \citep{vescovi20}. This mixing can cause a change in the metallicity dependence of the s-process production due to a change in the $\rm ^{13}C$ pocket in AGB stars, specifically for metal-rich stars in the inner regions. Therefore, the s-process yields including the mixing are lower than s-process yields without mixing. This scenario is a signature of the complexity of the s-process yields and, consequently, of the chemical evolution models of these elements. For this reason, a comparison between the observed age-chemical-composition trends and predictions from chemical evolution models can help us to understand the Ce production in our Galaxy.

Moreover, looking at the [Ce/$\alpha$]-age relations at different $R_{g}$, it is clear how their slopes become steeper moving towards outer regions (see Table~\ref{tab:mcmc}: $m_{1}$ becomes more negative with increasing $R_{g}$). This behaviour is a suggestion of the late enrichment of Ce in the Galaxy evolution due to the low- and intermediate-mass AGB stars and indicates a [Ce/$\alpha$] gradient in the Galactic disc. Moreover, there is a flattening at oldest age, a trend hinted at by \citet{salessilva22} for all studied $\alpha$-elements. However, we have the advantages of exploiting a homogeneous sample of stars with precise age, spanning the entire range in age and a large interval in metallicity. This is not possible using open clusters because they have ages typically younger than 7 Gyr and metallicity ranging [$-$0.4, +0.4] dex.  

To conclude, the [Ce/$\alpha$] trends with the stellar age, metallicity and $R_{g}$ are consistent in all data samples within $3\sigma$. The slight difference among the fitting parameters, particularly for K2, is likely due to the selection effects: our K2 targets have a broader $z$ distribution (where $z$ is the height from the Galactic mid-plane), with respect to \kepler and TESS targets. For a given $R_{g}$ bin, stars in K2 are located, on average, at higher distances from the Galactic plane. This different $z$ distribution implies a different distribution in age, since stars with higher $z$ are older. Indeed, a larger number of old stars in the K2 sample can change slightly the fitting parameters with respect to TESS and \kepler dataset.

\subsubsection{Application to field stars?}

Usually, the chemical clock relations obtained from a sample of stars with well-known ages (e.g., stars with asteroseismic age) are applied to other field stars of which we know the abundances, but we cannot derive age using standard techniques (e.g. isochrone fitting, asteroseismology etc.). There are some examples in literature, see for instance \citet{casali20} and \citet{viscasillas22} for the [s-/$\alpha$] ratios, but also \citet{casali19}, \citet{masseron15}, \citet{martig16} and \citet{ness16} for [C/N].

Unfortunately, the large spread in the Ce abundances does not permit to apply the relations in Table~\ref{tab:mcmc} to a sample of field stars. Indeed, at a given age, there are stars with a difference in [Ce/$\alpha$] of the order of $0.5$~dex, and the intrinsic scatter of these relations is of the order of $\sim 0.1$ dex. 
Part of this scatter can be ascribed to radial migration. Indeed, by using $R_{g}$ we can mitigate the scatter associated to the blurring effect only, not accounting for churning. To consider the full effect of radial migration, one would need to estimate the birth radii of stars (e.g. by interpolating the position of the stars in the age-metallicity relation in comparison with the chemical evolution curves for each radius from the models). In this way, the scatter would most likely be reduced. Nevertheless, this treatment is out of the scope of this work.

On the top of the expected scatter due to radial migration, there is also the fact that additional scatter is due to Ce measurement uncertainties. The uncertainties on the Ce abundances are likely underestimated as is the case for most of the uncertainties on other abundances and atmospheric parameters in APOGEE DR17. This is also proved by the large standard deviation in  the average abundances of  star  clusters in the Sect.~\ref{sec:ocs}. 
A better treatment of the Ce uncertainties might reduce the intrinsic scatter in the model and make these chemical clock relations useful to be applied to samples of field stars.
Indeed, we tested the sum in quadrature of the mean of the standard deviations measured for the open clusters shown in Fig.~\ref{fig:ocs} ($\sigma \sim 0.1$~dex) to Ce uncertainties. The result is a significant reduction of the intrinsic scatter ($\sim 50\%$) of the relations, while the fit parameters remain unaltered.

All these sources of scatter make impossible to apply the  [Ce/$\alpha$]-[Fe/H]-age relations to derive the ``chemical age'' of other field stars. 
Furthermore, the [Ce/$\alpha$]-age trends become flat for age $>$ 6 Gyr, so such chemical clock cannot be applied to old stars.



\begin{table*}[h]
 \caption{Parameters of the MCMC fitting for the [Ce/$\alpha$]-[Fe/H]-age relations in the {\it Kepler}, TESS and K2 samples. }
\begin{center}
\begin{tabular}{ccccccc}
\hline
Ratio & $m_{1}$~(dex/Gyr) &   $m_{2}$  &  $c$~(dex) & $n_{1}$~(dex/Gyr) & $k$~(Gyr) & $\epsilon$~(dex)  \\
\hline
& & & {\it Kepler} & & & \\
\hline
& & & $R_{g} < 7$~kpc & & & \\
\hline
$\rm [Ce/O]$  & $-0.013 \pm 0.002$ & $-0.074 \pm 0.036$ & $-0.136 \pm 0.015$ &  -- &  -- &  $0.135 \pm 0.005$  \\
$\rm [Ce/Mg]$ & $-0.011 \pm 0.002$ & $-0.132 \pm 0.037$ & $-0.147 \pm 0.015$ &  -- &  -- &  $0.136 \pm 0.005$  \\
$\rm [Ce/Si]$ & $-0.006 \pm 0.002$ & $-0.140 \pm 0.038$ & $-0.151 \pm 0.016$ &  -- &  -- &  $0.137 \pm 0.005$  \\
$\rm [Ce/Ca]$ & $-0.004 \pm 0.002$ & $-0.126 \pm 0.035$ & $-0.145 \pm 0.016$ &  -- &  -- &  $0.136 \pm 0.005$  \\
\hline
& & & $7 \leq R_{g} \leq 8$~kpc & & & \\
\hline
$\rm [Ce/O]$  & $-0.039 \pm 0.003$ & $-0.177 \pm 0.024$ & $0.038 \pm 0.013$ &  $-0.004 \pm 0.007$&  $6.1 \pm 0.6$ &  $0.086 \pm 0.004$  \\
$\rm [Ce/Mg]$ & $-0.041 \pm 0.004$ & $-0.211 \pm 0.025$ & $0.040 \pm 0.013$ &  $-0.002 \pm 0.007$&  $6.1 \pm 0.6$ &  $0.087 \pm 0.004$  \\
$\rm [Ce/Si]$ & $-0.034 \pm 0.003$ & $-0.229 \pm 0.023$ & $0.026 \pm 0.012$ &  $-0.001 \pm 0.005$&  $5.9 \pm 0.6$ &  $0.083 \pm 0.004$  \\
$\rm [Ce/Ca]$ & $-0.032 \pm 0.004$ & $-0.192 \pm 0.022$ & $0.027 \pm 0.013$ &  $ 0.000 \pm 0.005$&  $5.7 \pm 0.6$ &  $0.082 \pm 0.004$  \\
\hline
& & & $ R_{g} > 8$~kpc & & & \\
\hline
$\rm [Ce/O]$  & $-0.052 \pm 0.005$ & $-0.023 \pm 0.031$ & $0.103 \pm 0.015$ &  $ 0.002 \pm 0.020$ &  $6.9^{+1.0}_{-1.2}$ &  $0.096 \pm 0.005$  \\
$\rm [Ce/Mg]$ & $-0.050 \pm 0.004$ & $-0.049 \pm 0.029$ & $0.082 \pm 0.013$ &  $ 0.015 \pm 0.020$ &  $7.3^{+0.5}_{-1.2}$ &  $0.088 \pm 0.005$  \\
$\rm [Ce/Si]$ & $-0.042 \pm 0.004$ & $-0.067 \pm 0.028$ & $0.071 \pm 0.012$ &  $ 0.016 \pm 0.017$ &  $7.4^{+0.4}_{-0.9}$ &  $0.083 \pm 0.005$  \\
$\rm [Ce/Ca]$ & $-0.037 \pm 0.006$ & $-0.049 \pm 0.029$ & $0.057 \pm 0.017$ &  $ 0.010 \pm 0.019$ &  $7.2^{+0.7}_{-1.9}$ &  $0.083 \pm 0.005$  \\
\hline 
\hline 
 & & & TESS & & & \\
 \hline
& & & $R_{g} < 7$~kpc & & & \\
\hline
$\rm [Ce/O]$  & $-0.014 \pm 0.003$ & $-0.118 \pm 0.059$ & $-0.075 \pm 0.022$ &  -- &  -- &  $0.101 \pm 0.009$  \\
$\rm [Ce/Mg]$ & $-0.013 \pm 0.003$ & $-0.154 \pm 0.057$ & $-0.091 \pm 0.022$ &  -- &  -- &  $0.098 \pm 0.009$  \\
$\rm [Ce/Si]$ & $-0.011 \pm 0.003$ & $-0.163 \pm 0.055$ & $-0.077 \pm 0.021$ &  -- &  -- &  $0.095 \pm 0.008$  \\
$\rm [Ce/Ca]$ & $-0.008 \pm 0.003$ & $-0.154 \pm 0.055$ & $-0.064 \pm 0.022$ &  -- &  -- &  $0.097 \pm 0.008$  \\
\hline
& & & $7 \leq R_{g} \leq 8$~kpc & & & \\
\hline
$\rm [Ce/O]$  & $-0.035 \pm 0.004$ & $-0.056 \pm 0.042$ & $ 0.015 \pm 0.017$ &  $0.003 \pm 0.011$ &  $7.6^{+0.6}_{-0.7}$ &  $0.102 \pm 0.007$  \\
$\rm [Ce/Mg]$ & $-0.035 \pm 0.004$ & $-0.104 \pm 0.039$ & $ 0.008 \pm 0.017$ &  $0.006 \pm 0.011$ &  $7.7^{+0.6}_{-0.7}$ &  $0.099 \pm 0.006$  \\
$\rm [Ce/Si]$ & $-0.031 \pm 0.004$ & $-0.110 \pm 0.038$ & $-0.003 \pm 0.017$ &  $0.004 \pm 0.009$ &  $7.5 \pm 0.7$       &  $0.093 \pm 0.006$  \\
$\rm [Ce/Ca]$ & $-0.027 \pm 0.004$ & $-0.096 \pm 0.038$ & $ 0.013 \pm 0.016$ &  $0.006 \pm 0.010$ &  $7.6^{+0.7}_{-0.8}$ &  $0.088 \pm 0.006$  \\
\hline
& & & $ R_{g} > 8$~kpc & & & \\
\hline
$\rm [Ce/O]$  & $-0.040 \pm 0.004$ & $-0.013 \pm 0.044$ & $ 0.084 \pm 0.018$ &  $0.004 \pm 0.015$ &  $7.3^{+0.9}_{-1.1}$ &  $0.079 \pm 0.007$  \\
$\rm [Ce/Mg]$ & $-0.041 \pm 0.005$ & $-0.036 \pm 0.038$ & $ 0.085 \pm 0.018$ &  $0.001^{+0.013}_{-0.009}$ &  $6.8^{+0.9}_{-0.8}$ &  $0.077 \pm 0.007$  \\
$\rm [Ce/Si]$ & $-0.035 \pm 0.004$ & $-0.068 \pm 0.037$ & $ 0.065 \pm 0.017$ &  $0.002^{+0.011}_{-0.008}$ &  $6.7^{+1.6}_{-0.8}$ &  $0.068 \pm 0.007$  \\
$\rm [Ce/Ca]$ & $-0.030 \pm 0.004$ & $-0.044 \pm 0.033$ & $ 0.070 \pm 0.016$ &  $0.001 \pm 0.007$ &  $6.6^{+1.0}_{-0.8}$ &  $0.058 \pm 0.007$  \\
\hline
\hline 
 & & & K2 & & & \\
 \hline
& & & $R_{g} < 5$~kpc & & & \\
\hline
$\rm [Ce/O]$  & ${-0.006} \pm {0.005}$ & ${0.141} \pm {0.075}$ & ${-0.183} \pm {0.047}$ & -- & -- & ${0.113} \pm {0.017}$  \\
$\rm [Ce/Mg]$  & ${-0.004} \pm {0.005}$ & ${-0.017} \pm {0.075}$ & ${-0.181} \pm {0.047}$ & -- & -- & ${0.115} \pm {0.017}$  \\
$\rm [Ce/Si]$  & ${-0.001} \pm {0.005}$ & ${-0.045} \pm {0.074}$ & ${-0.182} \pm {0.047}$ & -- & -- & ${0.109} \pm {0.017}$  \\
$\rm [Ce/Ca]$  & ${-0.000} \pm {0.005}$ & ${0.059} \pm {0.074}$ & ${-0.171} \pm {0.048}$ & -- & -- & ${0.110} \pm {0.017}$  \\
\hline
& & & $5 \leq R_{g} < 7$~kpc & & & \\
\hline
$\rm [Ce/O]$  & ${-0.025} \pm {0.020}$ & ${-0.002} \pm {0.004}$ & ${-0.056} \pm {0.017}$& ${-0.004} \pm {0.005}$ & ${8.077} \pm {0.005}$ & ${0.091} \pm {0.026}$ \\
$\rm [Ce/Mg]$  & ${-0.026} \pm {0.020}$ & ${-0.031} \pm {0.004}$ & ${-0.059} \pm {0.017}$& ${-0.002} \pm {0.005}$ & ${7.863} \pm {0.005}$ & ${0.089} \pm {0.026}$ \\
$\rm [Ce/Si]$  & ${-0.019} \pm {0.020}$ & ${-0.072} \pm {0.004}$ & ${-0.066} \pm {0.017}$& ${-0.003} \pm {0.005}$ & ${7.434} \pm {0.004}$ & ${0.083} \pm {0.023}$ \\
$\rm [Ce/Ca]$  & ${-0.026} \pm {0.020}$ & ${-0.031} \pm {0.004}$ & ${-0.059} \pm {0.017}$& ${-0.002} \pm {0.006}$ & ${7.836} \pm {0.005}$ & ${0.089} \pm {0.025}$ \\
\hline
& & & $7 \leq R_{g} \leq 8$~kpc & & & \\
\hline
$\rm [Ce/O]$  & ${-0.020} \pm {0.003}$ & ${0.079} \pm {0.025}$ & ${-0.061} \pm {0.017}$& ${-0.007} \pm {0.005}$ & ${8.298} \pm {1.382}$ & ${0.114} \pm {0.005}$ \\
$\rm [Ce/Mg]$  & ${-0.021} \pm {0.003}$ & ${0.001} \pm {0.024}$ & ${-0.067} \pm {0.017}$& ${-0.008} \pm {0.005}$ & ${7.913} \pm {1.333}$ & ${0.109} \pm {0.004}$ \\
$\rm [Ce/Si]$  & ${-0.018} \pm {0.003}$ & ${-0.017} \pm {0.023}$ & ${-0.061} \pm {0.017}$& ${-0.006} \pm {0.004}$ & ${7.395} \pm {1.259}$ & ${0.101} \pm {0.004}$ \\
$\rm [Ce/Ca]$  & ${-0.021} \pm {0.003}$ & ${0.001} \pm {0.025}$ & ${-0.066} \pm {0.017}$& ${-0.008} \pm {0.005}$ & ${7.926} \pm {1.342}$ & ${0.109} \pm {0.004}$ \\
\hline
& & & $ 8 < R_{g} \leq 10$~kpc & & & \\
\hline
$\rm [Ce/O]$  & ${-0.035} \pm {0.002}$ & ${0.136} \pm {0.021}$ & ${0.047} \pm {0.011}$& ${0.004} \pm {0.007}$ & ${8.992} \pm {0.496}$ & ${0.111} \pm {0.004}$ \\
$\rm [Ce/Mg]$  & ${-0.034} \pm {0.002}$ & ${0.059} \pm {0.020}$ & ${0.034} \pm {0.011}$& ${0.004} \pm {0.006}$ & ${8.931} \pm {0.469}$ & ${0.103} \pm {0.004}$ \\
$\rm [Ce/Si]$  & ${-0.029} \pm {0.002}$ & ${0.018} \pm {0.019}$ & ${0.022} \pm {0.010}$& ${0.003} \pm {0.006}$ & ${8.812} \pm {0.553}$ & ${0.100} \pm {0.004}$ \\
$\rm [Ce/Ca]$  & ${-0.034} \pm {0.002}$ & ${0.061} \pm {0.019}$ & ${0.035} \pm {0.011}$& ${0.004} \pm {0.006}$ & ${8.916} \pm {0.482}$ & ${0.104} \pm {0.004}$ \\
\hline
& & & $R_{g} > 10 $~kpc & & & \\
\hline
$\rm [Ce/O]$  & ${-0.024} \pm {0.007}$ & ${0.204} \pm {0.057}$ & ${0.035} \pm {0.034}$& ${-0.018} \pm {0.011}$ & ${7.827} \pm {2.335}$ & ${0.125} \pm {0.009}$ \\
$\rm [Ce/Mg]$  & ${-0.024} \pm {0.006}$ & ${0.128} \pm {0.054}$ & ${0.029} \pm {0.031}$& ${-0.014} \pm {0.011}$ & ${8.172} \pm {2.082}$ & ${0.112} \pm {0.009}$ \\
$\rm [Ce/Si]$  & ${-0.021} \pm {0.006}$ & ${0.109} \pm {0.051}$ & ${0.033} \pm {0.030}$& ${-0.015} \pm {0.010}$ & ${7.881} \pm {2.316}$ & ${0.108} \pm {0.008}$ \\
$\rm [Ce/Ca]$  & ${-0.024} \pm {0.006}$ & ${0.128} \pm {0.054}$ & ${0.026} \pm {0.031}$& ${-0.014} \pm {0.012}$ & ${8.181} \pm {2.165}$ & ${0.112} \pm {0.009}$ \\
\hline 
\end{tabular}
\end{center}
\label{tab:mcmc}
\end{table*}


\section{Comparison with Galactic chemical evolution models}
\label{sec:models}

\begin{figure*}[ht]
\includegraphics[scale=0.45]{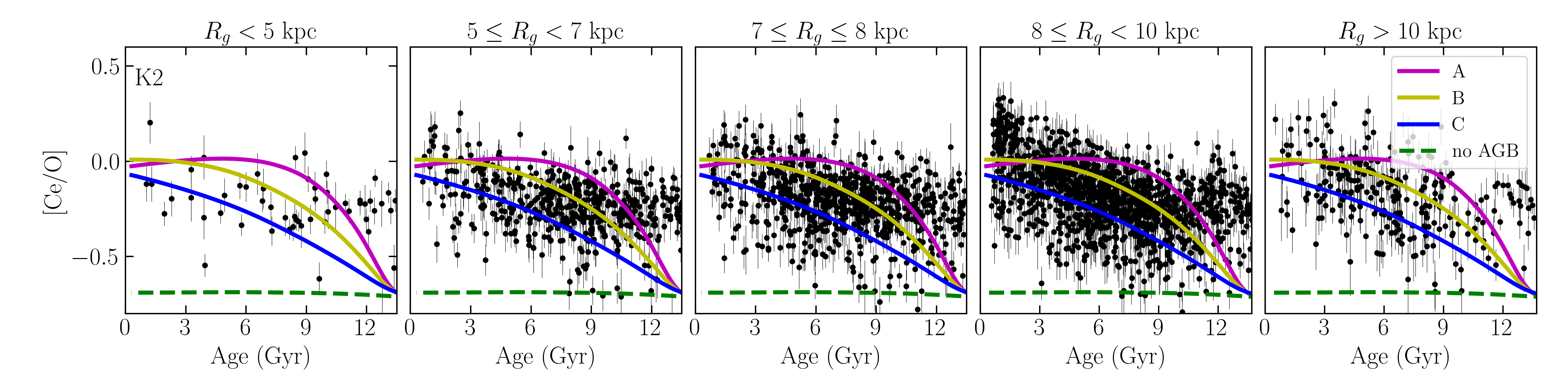}
\caption{Observed and predicted [Ce/O] vs age for the K2 sample in five bins of $R_{g}$. The Galactic chemical evolution models are labelled with the letters A, for the most efficient star formation history, B and C for the less efficient and the dashed line represents the GCE model without the AGB stars contributions. \label{fig:ceo_gce}}
\end{figure*}


Here, we  interpret our observational results in the light of chemical evolution models,  showing in particular the effect of different star formation efficiencies on the chemical evolution of Ce. A more quantitative analysis on Galactic chemical evolution (GCE) models and radial migration effects will be given in a future paper.

We compare the [Ce/$\alpha$] trends with the predictions of a GCE model. 
We focus on [Ce/O] ratio, since O is the most representative $\alpha$-element with the highest percentage of production from Type II Supernovae.

We consider the reference model of \citet{Grisoni2017,grisoni2020} to follow the evolution of the Galactic thin disc. In this model, Ce is produced by both the s- and r- process. Low- and intermediate-mass AGB stars in the range $1.3-6~\mathrm{M_{\odot}}$ are responsible for most of the s-process and the corresponding yields are taken from the database FRUITY \citep{cristallo09,cristallo11}. Moreover, we also include the s-process contribution from rotating massive stars, considering the nucleosynthesis prescriptions of \citet{Frischknecht2016} for these stars. The r-process yields for Ce are obtained by scaling the Eu yields
adopted in \citet{cescutti14}, according to the abundance ratios
observed in r-process rich stars \citep{Sneden2008}. For core-collapse supernovae, we consider the yields of \citet{kobayashi06}. 

In Fig.~\ref{fig:ceo_gce} we show the observed and predicted [Ce/O] vs age in different bins of $R_{g}$, where Ce and O have different timescales of production with Ce being mainly produced by low mass AGB stars and O by massive stars. The predictions are for the reference model of the Galactic thin disc \citep{grisoni2020}, with three different star formation histories: the reference case A (SFE $=0.5$~Gyr$^{-1}$), and also a less efficient star formation efficiency for model B (SFE $=0.2$~Gyr$^{-1}$) and model C (SFE $=0.1$~Gyr$^{-1}$). In order to better understand the fundamental contribution of AGB stars for the Ce production, we also show the model without AGB contribution. This shows a lower limit, where the contribution to the Ce abundance derives from the massive stars only. 
The chemical evolution of Ce is very sensitive to the star formation history, since the yields of Ce are very dependent on mass and metallicity. By assuming a more efficient star formation history (case A in Fig.~\ref{fig:ceo_gce}), we obtain a faster chemical evolution, reaching high values of [Ce/O] at earlier epochs 
and then the slope [Ce/O] vs age flattens. 
A milder star formation history, instead, shows an increase of [Ce/O] towards younger ages, due to the fact that AGB stars contribute at later times with respect to the massive ones. 

Because of the large scatter due to the uncertainties on the Ce measurements and  to stellar radial migration, the three models A, B, and C can only be used to describe the trends of our data in each bin of guiding radius.
Model A represents better the [Ce/O] trend in the inner regions, while the models with a milder SFE (model B and C) describes the [Ce/O] behaviour in the outer regions, as expected also from inside-out scenario \citep{matteucci89,chiappini01,Grisoni2018,spitoni2021}. 
In fact, in agreement with the inside-out, we assume that the inner Galactic regions form faster than the outer ones: in this way, the larger number of stars per unit time in the inner regions allows to reach earlier the maximum [Ce/O] value with respect to the outer regions, which evolve with a milder SFE.
To conclude, the inside-out scenario, combined with the metallicity dependence of AGB yields, can explain the findings discussed in the previous sections. 

Moreover, the increasing trends of the observed [Ce/O] moving towards the outer regions is a possible signature of a [Ce/O] gradient in the Galactic disc. This gradient is made less clear because of the large scatter of the data. The reason for this scatter could be related to the uncertainties of the Ce measurements and radial migration. Even though we used the $R_{g}$ to mitigate the blurring effect, there is still the churning component which makes the study of abundance gradients in the Galactic disc very complicated. The treatment of the radial migration is beyond the scope of this work, but 
we can overcome this problem by focusing on the youngest ages ($< 1$~Gyr). Indeed, it is expected that radial migration
does not affect (or at least only slightly) young stars, because they have not had time to migrate yet.

\begin{figure}
\includegraphics[scale=0.47]{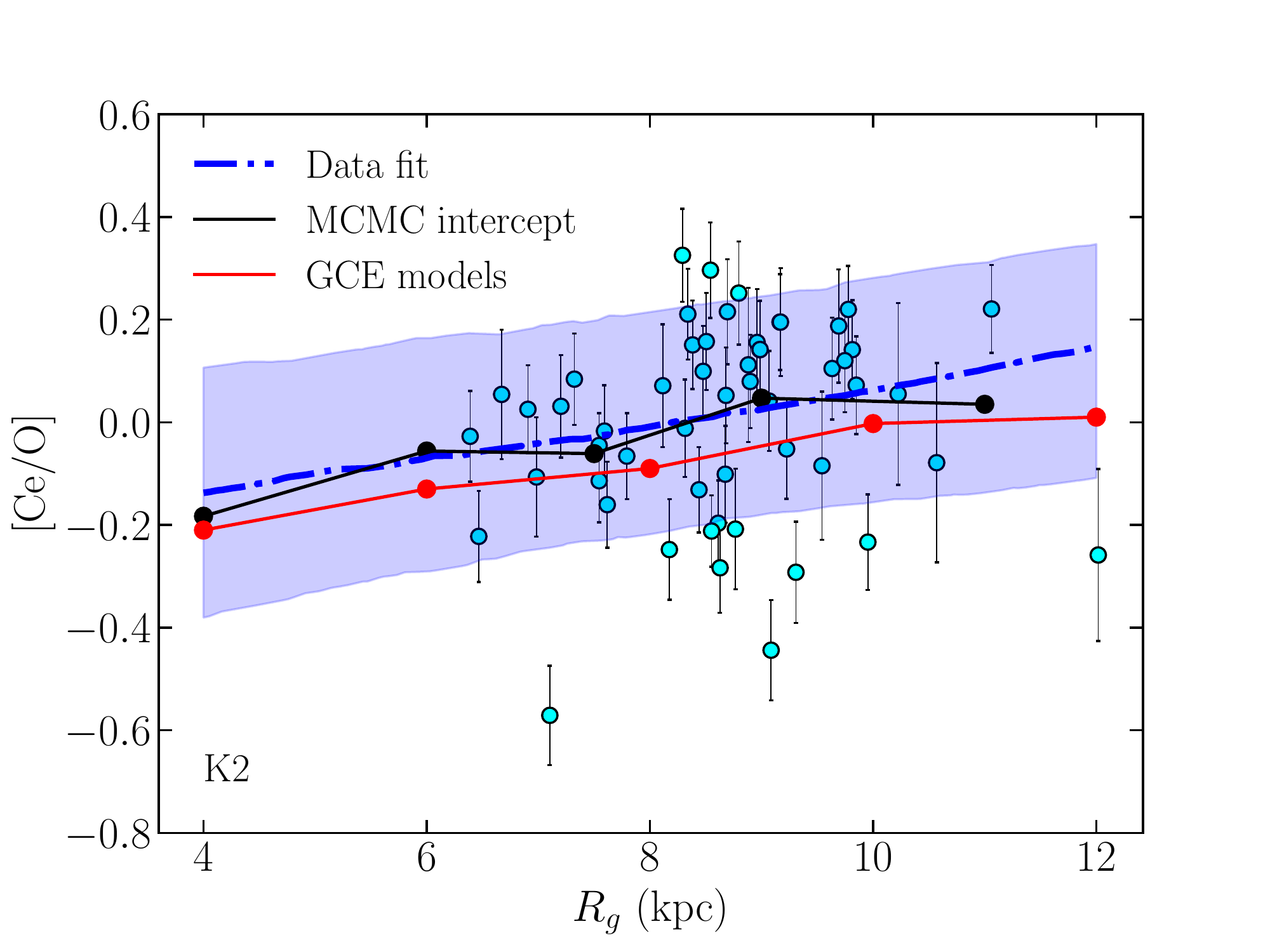}
\caption{[Ce/O] gradient for K2 stars with age $<$ 1 Gyr (cyan circles). The blue dashed line represents the best fit and the blue shaded area the 68\% confidence interval plus intrinsic scatter; the black line the intercept of the MCMC simulation studied in Table~\ref{tab:mcmc} at the present time in the five $R_{g}$ bins; the red line the GCE models at the present time at 4, 6, 8, 10 and 12 kpc.
\label{fig:grad}}
\end{figure}

In Fig.~\ref{fig:grad}, we plot the K2 stars with age $< 1$~Gyr (the age bin less affected by radial migration)\footnote{See \citet{anders17} who found the metallicity gradients for CoRoT stars younger than 1~Gyr to be similar to the one traced by Cepheids and HII regions}. 
There is still scatter in this age bin, but we can see a [Ce/O] gradient in the data. The fit is shown in the figure and is compared with the points at the present time of the GCE model at different Galactocentric distances \citep{Grisoni2018} and the intercepts of the MCMC models shown in Table~\ref{tab:mcmc} at different bins of radii. 
The linear regression is obtained using a MCMC simulation.
We can see a good agreement between the fit of the data with the gradient predicted by models at the present time.  Nevertheless, the spread of the linear regression is very large.

Our goal is neither to give a robust theoretical explanation of these trends nor to study in detail the gradients at this stage, but to present the results from an observational point of view. 

\section{Summary and Conclusions}
\label{sec:conclusions}
The study of the abundance trends of s-process elements in our Galaxy is very important for constraining theoretical models. Thanks to the data collected by the Gaia mission,  by large spectroscopic surveys, such as Gaia-ESO, GALAH and APOGEE, and space missions, such as {\it Kepler}, TESS, K2 and CoRoT, we can start exploiting the orthogonal constraints offered by age and chemistry to infer s-process elements history, and consequently, the Milky Way formation and evolution.

In this work we focus on the heavy s-process element, Ce, for three different datasets: {\it Kepler}, TESS SCVZ and K2. 
For all of them, we have abundances from the APOGEE DR17, where Ce is the only measured s-process element in this survey. Currently, APOGEE is the only high-resolution large-spectroscopic survey that can be cross-matched with the TESS and {\it Kepler} dataset.

The TESS and {\it Kepler} samples observed stars in the solar neighbourhood, while K2 observed different regions of our Galaxy, spanning a large range in Galactocentric distance. 
Furthermore, using these three datasets of field stars, we are able to cover the oldest age, not covered by open clusters \citep[see, e.g.,][]{salessilva22,casamiquela21,viscasillas22} and benefit from the improved statistics of a larger sample. Indeed, open clusters can cover ages up to 7 Gyr, with the bulk of them around 1 Gyr. Finally, the number of observed (spectroscopically) open clusters is less than $\sim 150$ \citep[e.g., see open clusters in the Gaia-ESO survey, OCCAM, OCCASO,][]{randich22,myers22,occaso}.

In this study, we investigate the [Ce/Fe] abundance ratio trends and the temporal evolution of Ce. Here are our main findings:
\begin{itemize}
\item In the [Ce/Fe] vs. [$\alpha$/M] plane, we see how the two population with high- and low-$\alpha$ content have different ages and a different average [Ce/Fe]. The high-$\alpha$ sequence contains older stars with lower mean [Ce/Fe], while the low-$\alpha$ sequence displays a large range in stellar age with higher mean [Ce/Fe]. Hereafter, we focus on the low-$\alpha$ sequence only, since the low- and intermediate-mass AGB stars polluted mainly this stellar population.

\item In the [Ce/Fe] vs. [Fe/H] plane, we see a peak in [Ce/Fe] ratio at around $-0.2$ dex in [Fe/H] with a consequent decrease of [Ce/Fe] towards lower and higher metallicity. This is related to a non-monotonic metallicity dependence of the s-process stellar yields, and a decreasing efficiency of the s-process at high metallicity. Moreover, at a given [Fe/H], younger stars have a higher [Ce/Fe] content than the older stars. A similar behavoiur is studied in open clusters by, e.g., \citet{salessilva22} and \citet{viscasillas22}. 

\item To study the time evolution of Ce, we 
divide the samples in different bins of guiding radius $ R_{g}$. The [Ce/Fe] vs. age plane shows how metal-poor stars have a content of [Ce/Fe] larger than metal-rich stars at a given age in all bin of $R_{g}$. 
Stars with $R_{g} < 7$ kpc show a flattening trend with age. This is a possible signature of the high star formation efficiency in the inner regions. 
In the other bins of $R_{g}$ ($R_{g} > 7$~kpc), [Ce/Fe] decreases with increasing age with a different slope. Moreover, for ages $< 6-8$~Gyr the trend is steeper, while for ages $> 6-8$~Gyr the trend is almost flat, confirming the results for open clusters in \citet{salessilva22}. This is a possible signature of the latest enrichment of low-mass AGB stars for Ce, i.e. inside out scenario, combined with metal-dependent AGB yields.  
This late enrichment takes a few Gyr to pollute the interstellar medium. Furthermore, the [Ce/Fe]-age slope becomes steeper and steeper moving towards the outer regions. 

\item Then, we study the [Ce/$\alpha$] as a chemical clock for the low-$\alpha$ sequence in the same bins of $R_{g}$. The ratio between Ce and an $\alpha$ element maximises the correlation with the stellar age. The trend is decreasing with age, showing that younger stars have a higher content of [Ce/$\alpha$] than the older ones. Moreover, we can see a different behaviour at different metallicity, as we already saw in literature, and a change in slope at 6-8 Gyr for these ratios. 
As we have already seen for [Ce/Fe]-age, also [Ce/$\alpha$]-age trends show a steeper slope in the young regime moving towards the outer disc.
However, the large scatter due to the underestimate of the Ce uncertainties and radial migration does not allow us to apply the [Ce/$\alpha$]-age-[Fe/H] relations to date stars for which we cannot derive age with other stellar age dating methods. 

\item Finally, we compare the [Ce/O]-age relations with the predictions of Galactic chemical evolution models. This comparison supports the interpretation that the [Ce/O] -- but also [Ce/$\alpha$] in general -- trends with the $R_{g}$ seen in the observational data are indeed related to the evolution of the Galactic disc: the observed behaviour is in agreement with the metallicity dependence of the AGB yields and inside-out scenario, where the inner parts of the Galaxy form faster than the outer ones. 
Indeed, GCE models with a more intense SFE represents better the observed [Ce/O] trend in the inner regions, while models with a less efficient SFE describes the [Ce/O] behaviour in the outer regions.
These GCE models allows us to reproduce also the present-day [Ce/O] gradient.
\end{itemize}

To conclude, our results show a strong dependence of Ce on metallicity, stellar ages and position in the Galactic disc.  
Moreover, Ce, a heavy s-process element, is mainly produced by low-mass AGB stars. Their longer lifetime implies a delayed ejection with respect to intermediate and massive stars, and a late contribution to the Galactic chemical evolution.
Finally, the uncertainties on s-process yields due to the metallicity dependence, treatment of the $\rm ^{13} C$-pocket, convection, mass-loss rates, etc. require  further investigations, which are likely to shed light on the trends of Ce abundances -- and those of the other s-process elements -- with age. 


\section*{Acknowledgements}
GC, VG, AM, MM, EW, AS, MM, JS acknowledge
support from the European Research Council Consolidator Grant funding scheme (project ASTEROCHRONOMETRY, G.A. n. 772293, \url{http://www.asterochronometry.eu}. The authors thank Thomas Masseron to providing us with the BAWLAS catalogue for the comparison. 
LM and GC acknowledge support from INAF with the Grant "Checs, (CHEmical ClockS) Seeking a theoretical foundation for the use of chemical clocks".

\bibliographystyle{aa}
\bibliography{Bibliography}

\begin{appendix}
\section{Mock data}
\label{app:mock}
We generate mock data using a uniform distribution for [Fe/H], between $-0.5$ and 0.3 dex, and a truncated Gaussian for the age distribution, with a median of 7 Gyr, bound between 1 and 12 Gyr, and a $\sigma$ of 3.5 Gyr.
These choices  aim at reproducing broadly the observed distributions.

\begin{figure}[ht]
\centering
\includegraphics[scale=0.4]{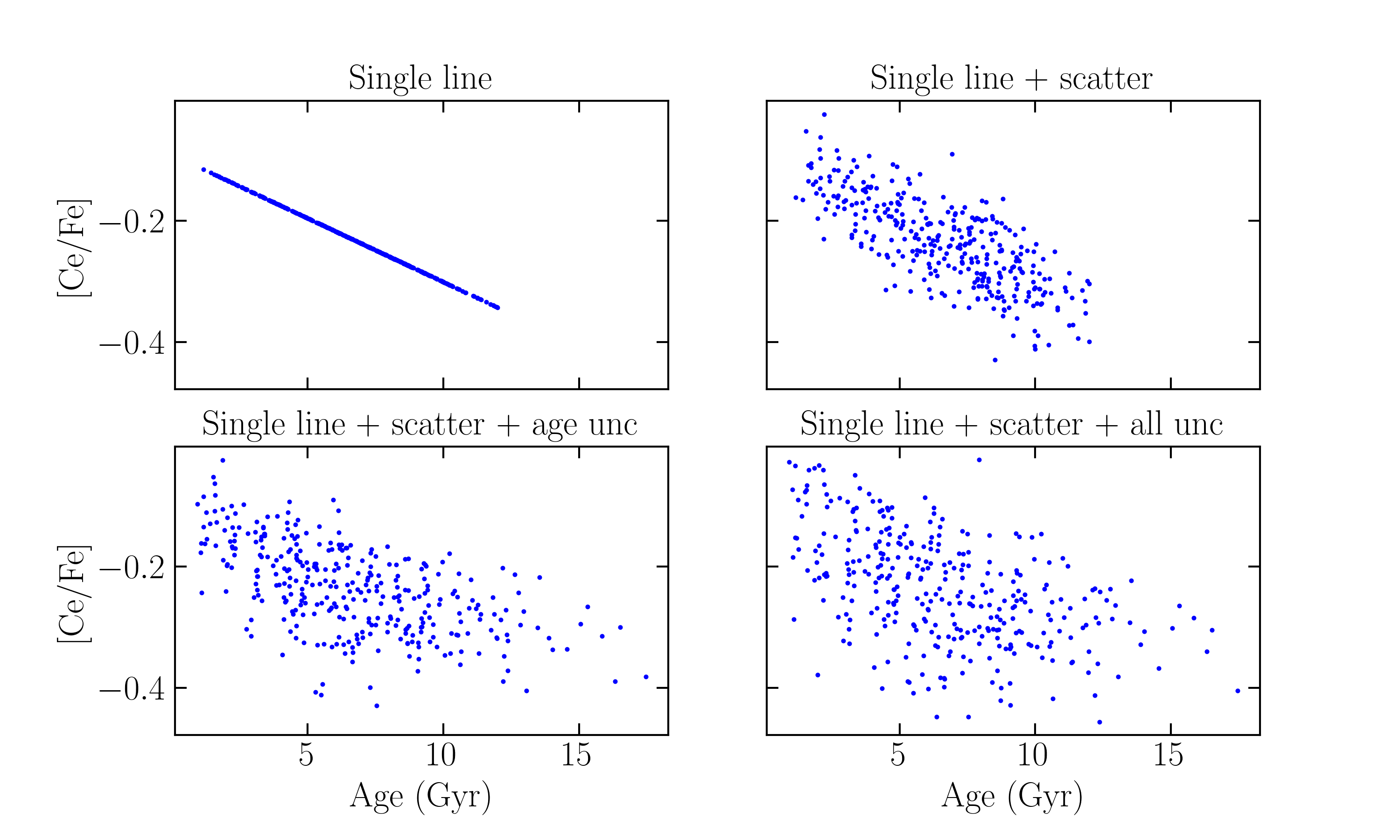}
\caption{The distribution of the mock data using a single-line model. The second panel at the top is the model including the intrinsic scatter, the first panel at the bottom is the model with intrinsic scatter and age uncertainties, the second one at the bottom is the model including the intrinsic scatter and all uncertainties. \label{fig:mockdata}}
\end{figure}

Then, we define a function to generate [Ce/Fe] based on a single-line model (as explained in Sect.~\ref{sec:cefe_age}). Then, we add a scatter and, afterwards, the uncertainties on [Fe/H], [Ce/Fe] and relative age: $\sigma_{\rm [Fe/H]}=0.05$~dex;
$\sigma_{\rm [Ce/Fe]}=0.05$~dex; $\sigma_{Age}=30\%$ (see Fig.~\ref{fig:mockdata}).

Finally, we perform the MCMC simulations applying the broken-line and the single-line model to the data (with scatter and uncertainties included) generated using a single-line. We do this to understand which of the two models can reproduce the mock data better. If the broken-line reproduces the mock data better, this means the change in slope in the [Ce/Fe]-age plane that we see in our data is due to the larger age uncertainties of older stars. If, instead, the single-line reproduces the observational data better, this implies the change in slope might be due to the Galactic evolution.

To discern which model fits better the mock data, we use the Bayesian Information Criterion (BIC) parameter. The model with the lowest BIC is the most reasonable one. 
We obtain a BIC of $-803.13$ for the broken-line  
and of $-813.68$ for the single-line. 
Following the rules by \citet{bayesfactors}, we can compute $\rm \Delta BIC = BIC_{i} - BIC_{min}$. Given $i$ models, the magnitude of the $\rm \Delta BIC$ can be interpreted as evidence against a candidate model being the best model. The
rules of thumb are: (i) less than 2, it is barely worth mentioning; (ii) between 2 and 6, the evidence against the candidate model is positive; (iii) between 6 and 10, the evidence against the candidate model is strong; (iv) greater than 10, the evidence is very strong.
In our case, the $\rm \Delta BIC$ is larger than 10.
This shows that the single-line model is the most reasonable one.


%
%

\section{Additional figures}
\begin{figure*}[ht]
\includegraphics[scale=0.3]{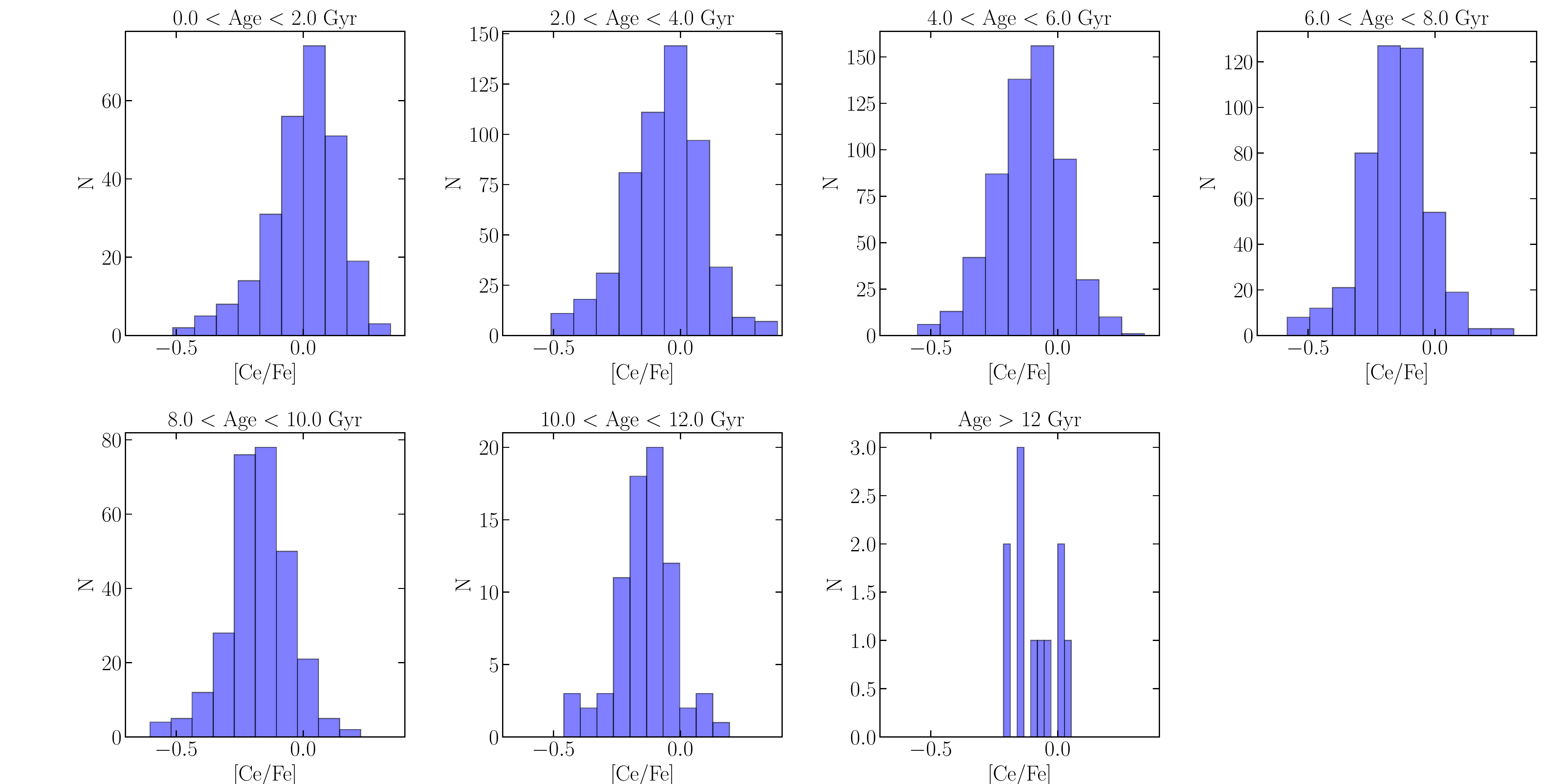}
\includegraphics[scale=0.3]{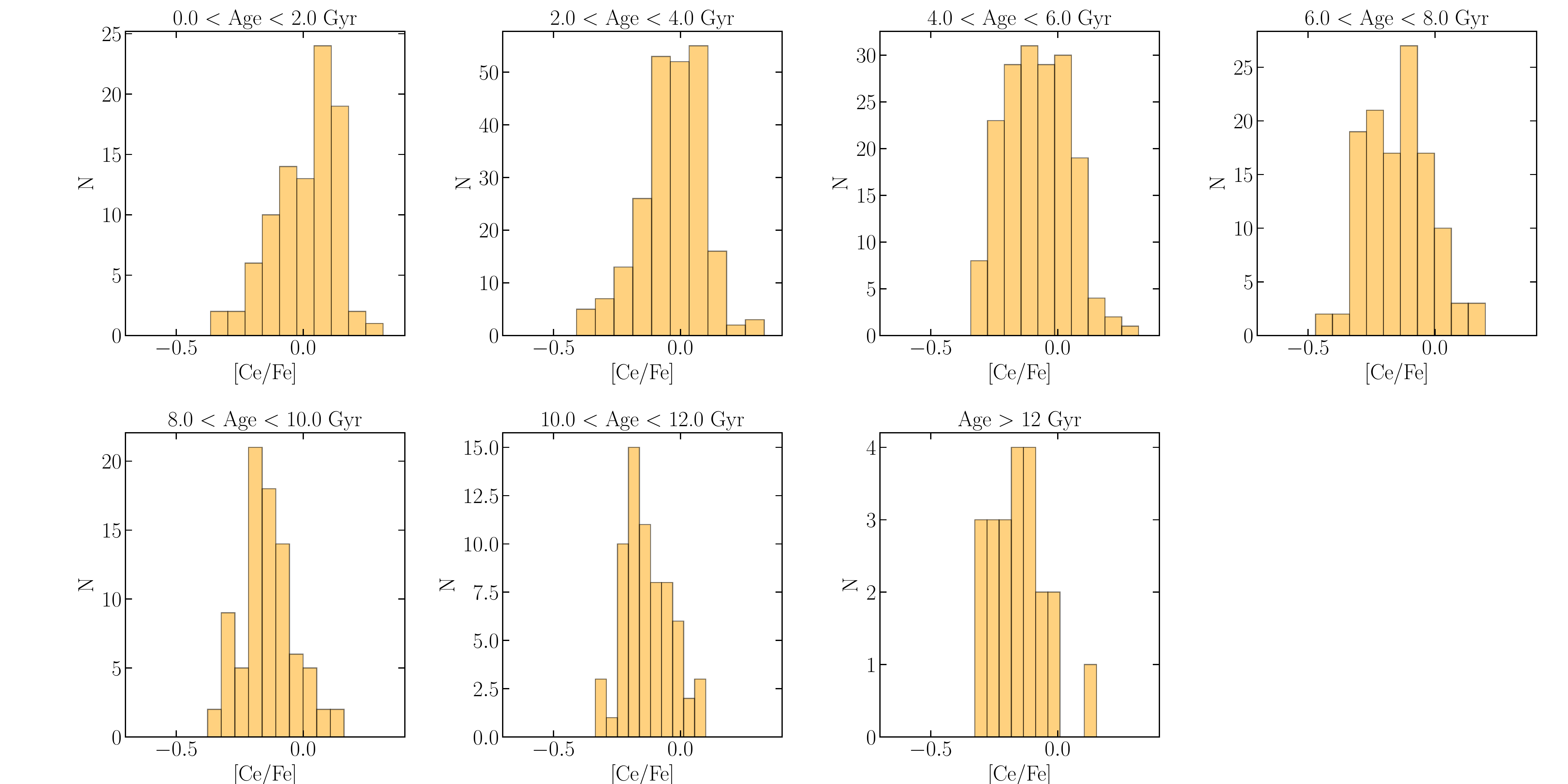}
\includegraphics[scale=0.3]{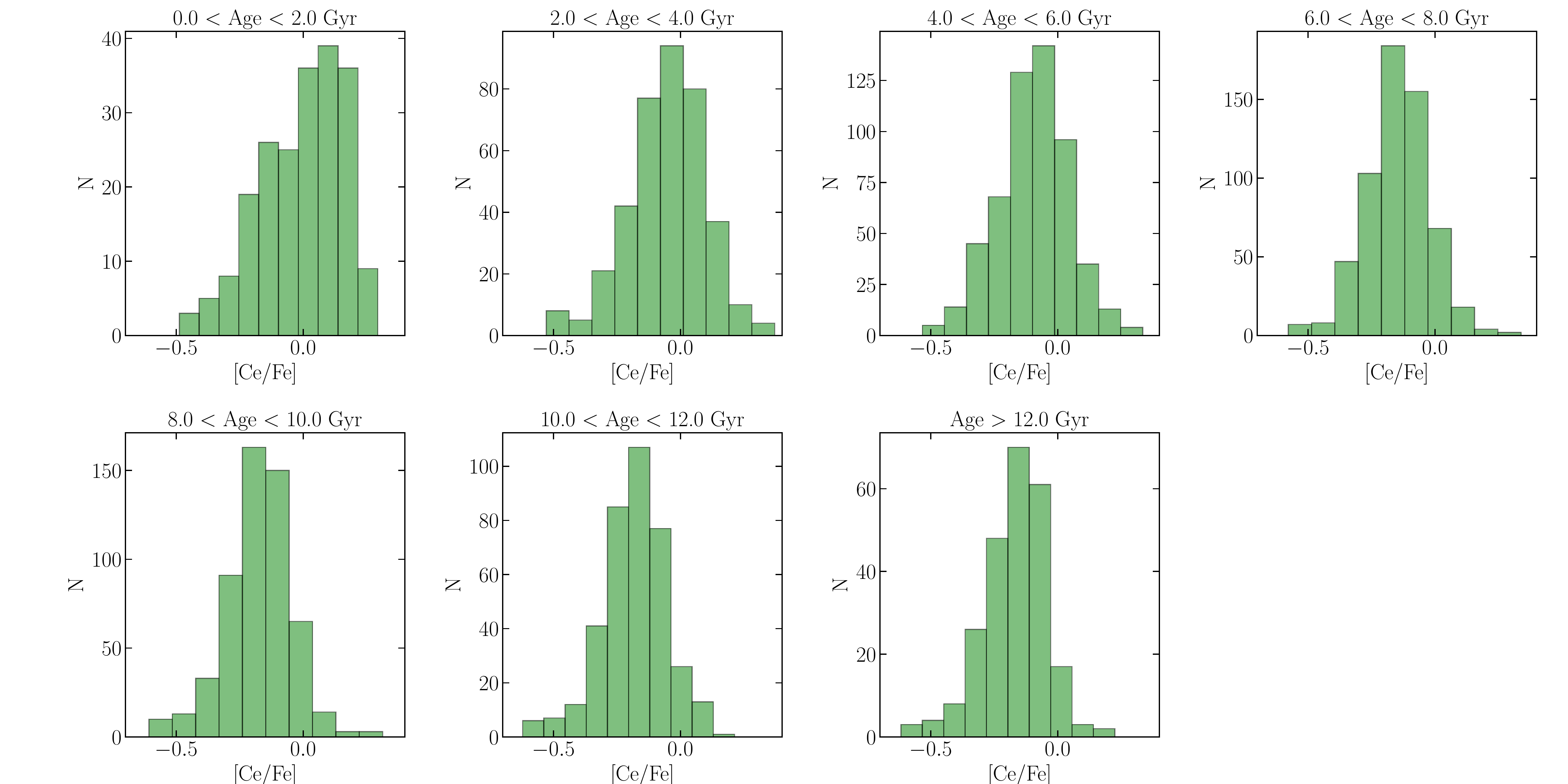}
\caption{The [Ce/Fe] distributions of the low-$\alpha$ sequences of {\it Kepler} (top, blue), TESS (middle, orange) and K2 (bottom, green) in different bins of stellar age.  \label{fig:hist}}
\end{figure*}

\begin{figure*}[ht]
\centering
\includegraphics[scale=0.45]{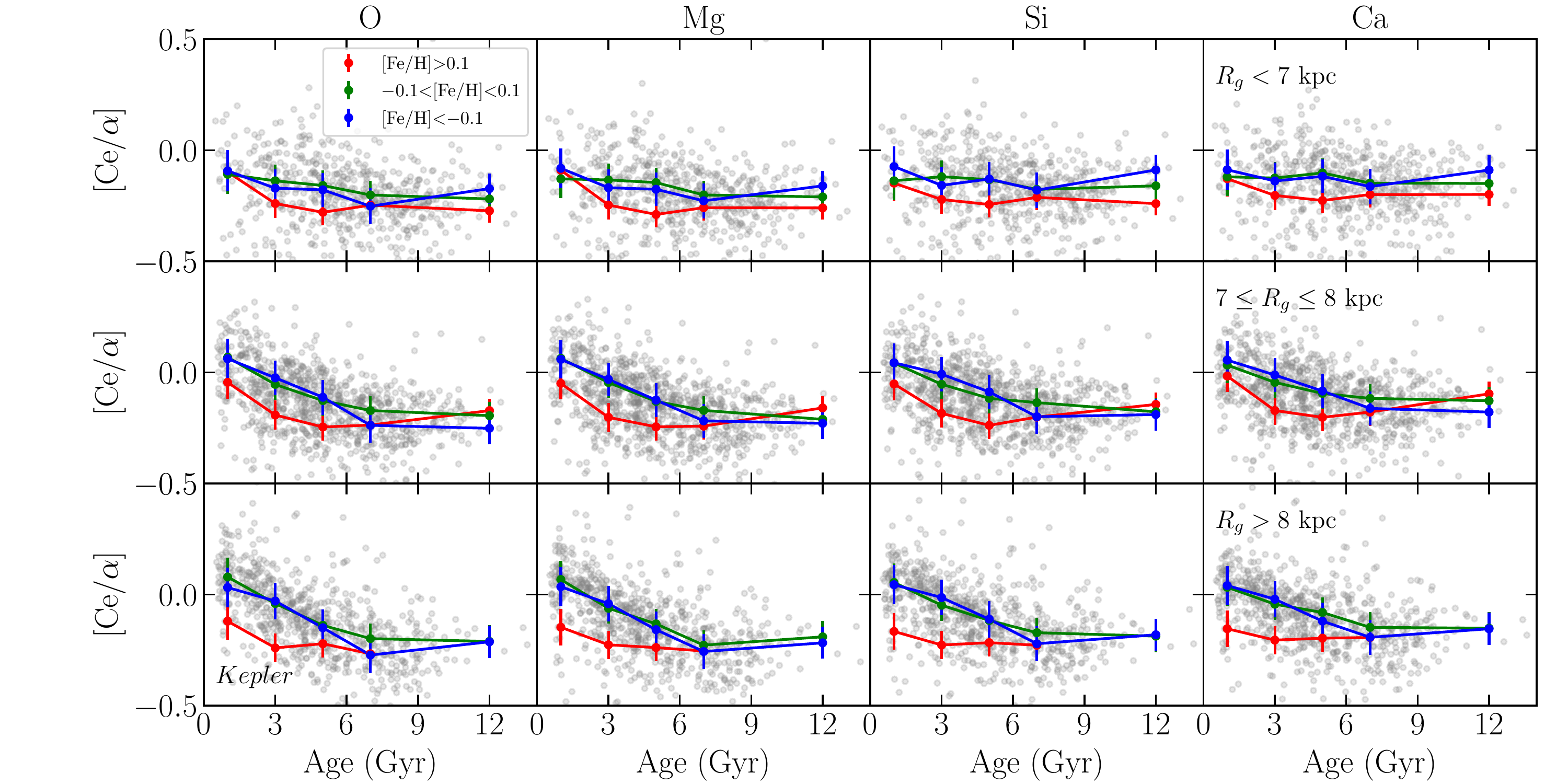}
\includegraphics[scale=0.45]{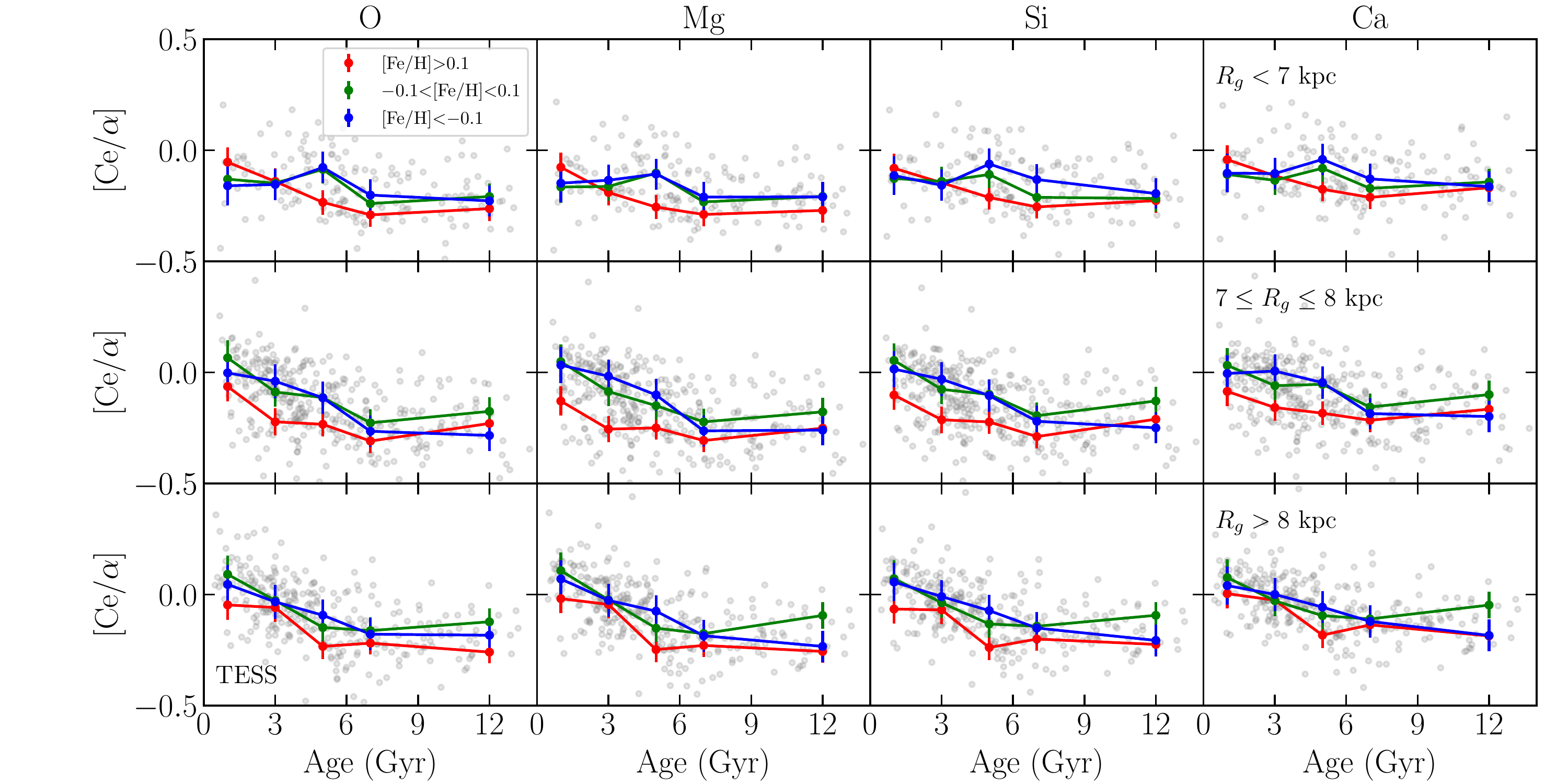}
\includegraphics[scale=0.45]{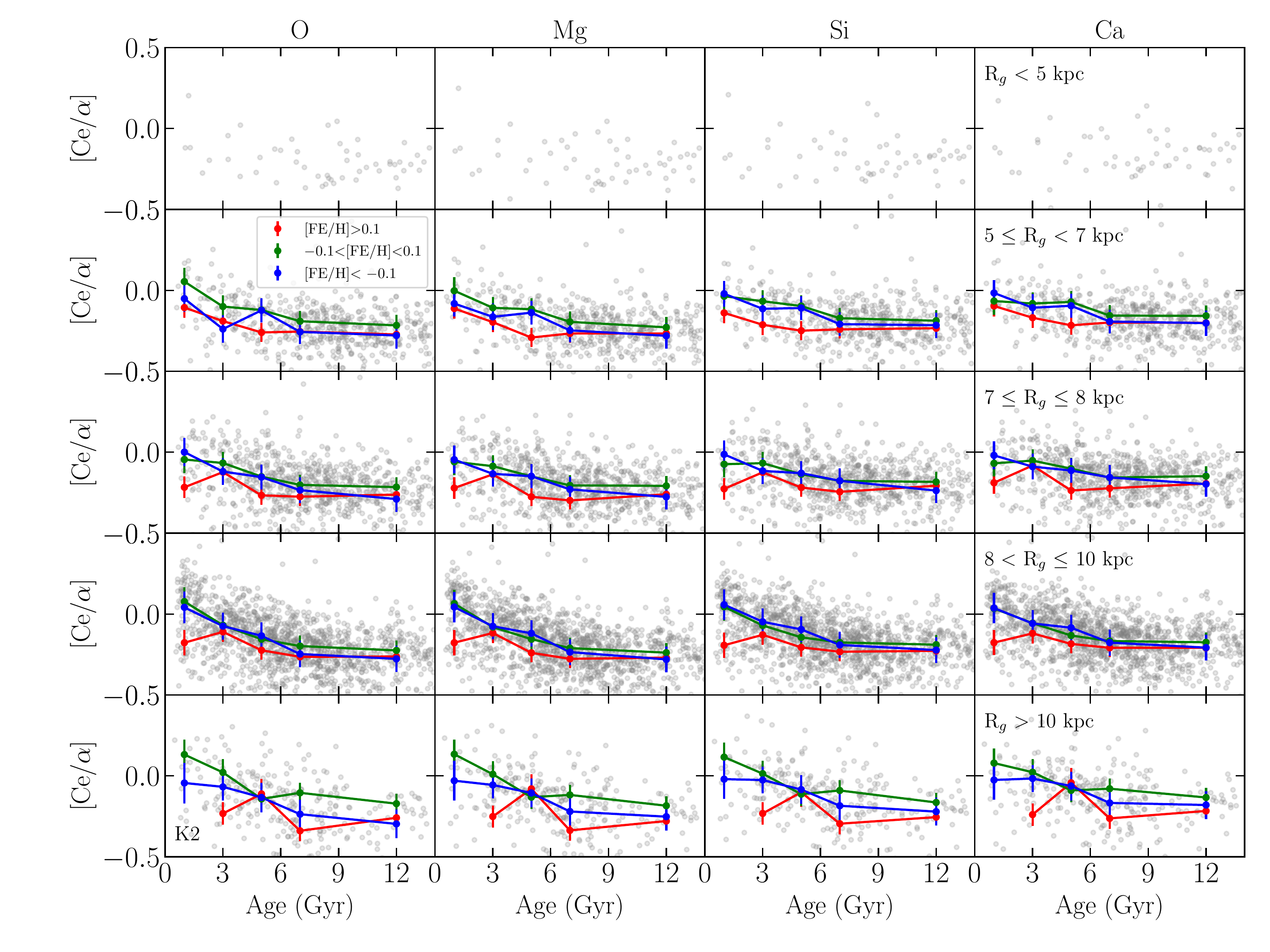}
\caption{[Ce/$\alpha$]-age planes for the low-$\alpha$  {\it Kepler} (top) and TESS (middle) and K2 (bottom) stars in different bins of $R_{g}$. The three lines represent the [Ce/$\alpha$] average in three different bins of metallicity: $\rm  [Fe/H] < -0.1$, $\rm  -0.1 \leq [Fe/H] \leq +0.1$, $\rm [Fe/H] > +0.1$. \label{fig:CeX_fehbins}}
\end{figure*}

In this appendix, we find some additional figures to a better understanding of our work. 
The Fig.~\ref{fig:hist} displays the [Ce/Fe] uncertainties distributions for the {\it Kepler}, TESS and K2 stars in different bins of age, while Fig.~\ref{fig:CeX_fehbins} shows the [Ce/$\alpha$]-age planes for the three data samples in different bins of $R_{g}$. The three different lines in Fig.~\ref{fig:CeX_fehbins} represent the [Ce/$\alpha$] average in three different bins of metallicity: $\rm  [Fe/H] < -0.1$, $\rm  -0.1 \leq [Fe/H] \leq +0.1$, $\rm [Fe/H] > +0.1$. For the K2 stars in the interval $R_{g} < 5$~kpc, we do not plot the [Ce/$\alpha$] average because there are too few stars. 

\end{appendix}

\end{document}